\documentclass[a4paper]{iopart}
\usepackage[T1]{fontenc}
\usepackage[utf8]{inputenc}
\usepackage{float,graphicx}
\makeatletter
\usepackage{iopams,setstack}
\makeatother
\usepackage[english]{babel}

\usepackage[usenames,dvipsnames]{color}
\definecolor{mygreen}{rgb}{0.2,0.6,0}
\definecolor{myblue}{rgb}{0,0,0.75}
\definecolor{mymagenta}{cmyk}{0,1,0,0.12}

\newcommand{\iu}{{\mathrm{i}}}
\usepackage[colorlinks,bookmarks=false,citecolor=blue,linkcolor=red,urlcolor=blue]{hyperref}
\usepackage{babel}
\begin{document}

\title[Dynamical preparation of laser-excited anisotropic Rydberg crystals]{Dynamical preparation of laser-excited anisotropic Rydberg crystals in 2D optical lattices}
\author{B.~Vermersch$^{1,2}$, M.~Punk$^{1,2}$, A.~W.~Glaetzle$^{1,2}$, C.~Gross$^3$ and P.~Zoller$^{1,2}$}
\address{$^{1}$ Institute for Quantum Optics and
Quantum Information, Austrian Academy of Sciences, 6020 Innsbruck, Austria}
\address{$^{2}$ Institute for Theoretical Physics,  University of Innsbruck,
6020 Innsbruck, Austria}
\address{$^{3}$ Max-Planck-Institut f\"{u}r Quantenoptik, 85748 Garching, Germany}
\date{\today}

\begin{abstract} We describe the dynamical preparation of anisotropic crystalline phases obtained by laser-exciting ultracold Alkali atoms to Rydberg p-states where they interact via anisotropic van der Waals interactions. We develop a time-dependent variational mean field ansatz to model large, but finite two-dimensional systems in  experimentally accessible parameter regimes, and we present numerical simulations to illustrate the dynamical formation of anisotropic Rydberg crystals. 
\end{abstract}
\submitto{\NJP}
\maketitle

\section{Introduction}

Highly excited Rydberg states of atoms have unique properties. This includes the size of the Rydberg orbitals scaling as $n^2$, the polarizabilities as $n^7$ and a long lifetime as $n^3$ with $n$ the principal quantum number. These properties are also manifest in interactions between Rydberg states, e.g.~in van der Waals (vdW) interactions $\propto n^{11}/r^6$, which can be controlled and tuned via external fields. Exciting ground state atoms with a laser to Rydberg states thus provides a means to study many body systems with strong, long-range interactions \cite{Saffman2010,Low2012}. With the atomic ground state and the Rydberg state defining an effective spin-$1/2$, we can describe the many body dynamics in terms of a model of interacting spins \cite{Bloch2008,Bloch2012,Georgescu2014}, reflecting the competition between the laser excitation and vdW interactions, at least in the short time limit where the motion of the atoms can be neglected (the so-called frozen gas regime). 

The study of quantum phases of a laser excited Rydberg gas of alkali atoms, including its dynamical preparation, has so far focused on isotropic  vdW interactions, corresponding to Rydberg s-states excited in a two-photon process. This includes theoretical studies  \cite{Robicheaux2005,Weimer2008,Pohl2010,Schachenmayer2010,Lesanovsky2011,Garttner2012} and experimental observations  \cite{Schauss2012,Schauss2014} of Rydberg crystals due to the Rydberg blockade mechanism \cite{Jaksch2000,Lukin2001,Tong2004,Singer2004,Cheinet2008,Gaetan2009,Urban2009}, and their melting with increasing laser intensity to a quantum-disordered phase \cite{Weimer2010,Sela2011}. The steady state of the system has also been studied in presence of dissipation \cite{Honing2013,Lesanovsky2013,Petrosyan2013,Hoening}. With the availability of UV laser sources also Rydberg p-states can be excited in a single photon transition, leading to anisotropic vdW interactions \cite{Reinhard2007,Glaetzle2014}. The goal of this paper is to investigate the quantum phases and their dynamical preparation with a laser pulse protocol for these anisotropic interactions. We are interested in 2D systems with a relatively high density of excitations involving a larger number of atoms, and in particular in the dynamical formation of anisotropic Rydberg crystals. Our studies are performed within a time-dependent variational mean field ansatz, beyond what can be accessed by exact diagonalization techniques.

\section{Model and Method\label{sec:model}}

\subsection{Laser excited interacting Rydberg atoms as an anisotropic spin model}

\begin{figure}
\begin{centering}
\includegraphics[width=0.95\columnwidth]{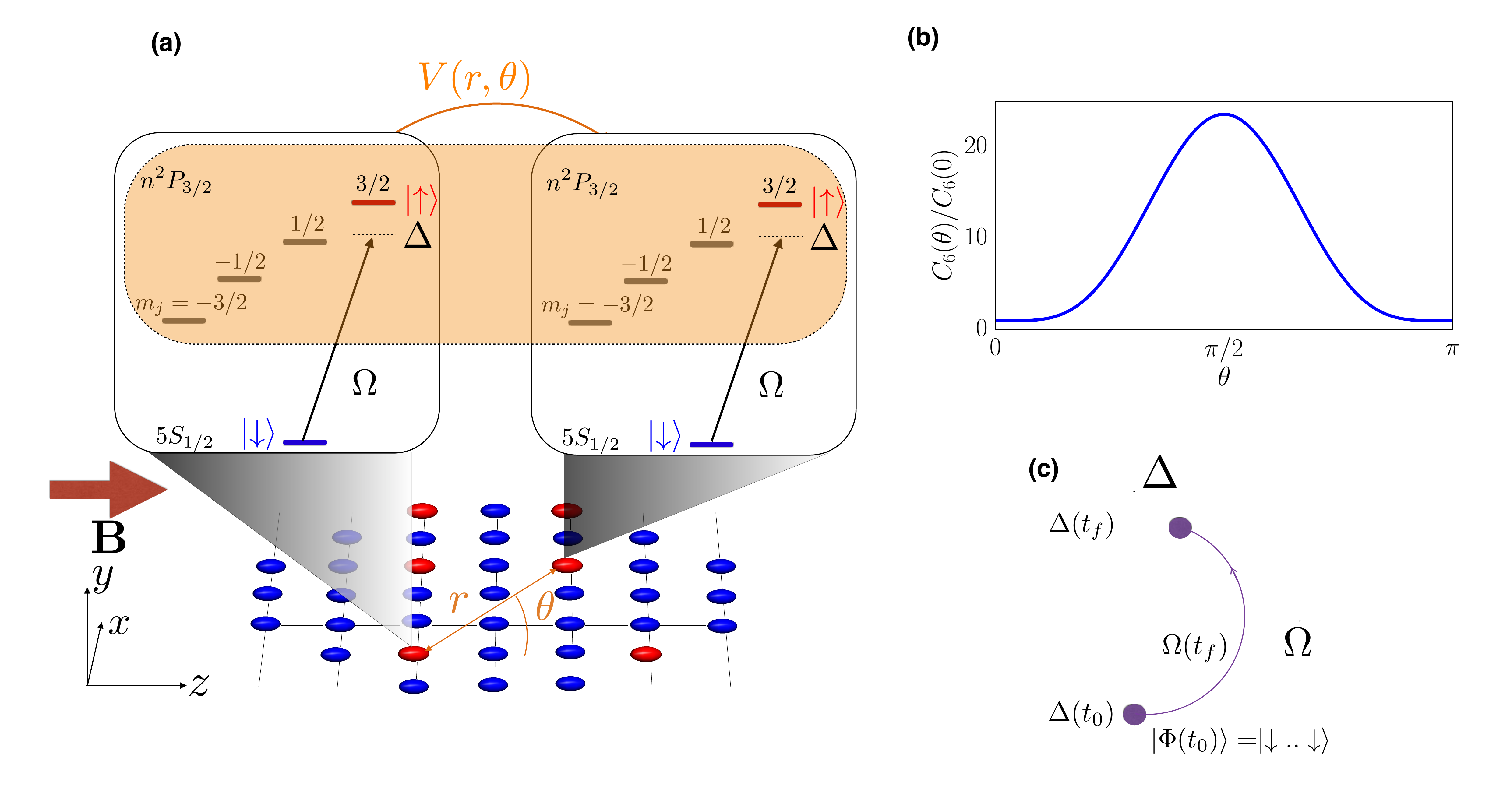}
\protect\caption{(a) Setup: The ground state atoms $\mid\downarrow\rangle$ are placed in a square
optical lattice and are excited to a Rydberg state $\mid\uparrow\rangle$
via a homogeneous laser beam with Rabi frequency $\Omega$ and detuning
$\Delta$. The vdW interaction $V$ between two Rydberg atoms $i$
and $k$ is a function of their relative distance $|\mathbf{r}_{i}-\mathbf{r}_{k}|$
but also of the angle $\theta_{i,k}$ between their relative vector
and the magnetic field $B$ which is set along the $z$ direction of the lattice. The details of these interactions in the fine structure manifold $n^2P_{3/2}$, in the presence of the magnetic field are discussed in \sref{sec:atomic}.  (b) Example of angular dependence of the $C_6$ coefficient obtained for a Rydberg state of Rubidium $|25P_{3/2},m_j=3/2\rangle$. (c) Example of sweep path: initially the atoms are prepared in the ground state $\mid\downarrow\rangle$ with a negative detuning $\Delta(t_0)<0$. The Rabi frequency $\Omega$ and the laser detuning $\Delta$ are then slowly varied to reach the final state of the preparation at time $t_f$. \label{fig:setup}}
\end{centering}
\end{figure}

We are interested in the quantum dynamics and the quantum phases of
a gas of laser excited Rydberg atoms, interacting via \emph{anisotropic}
vdW interactions. The setup we have in mind is represented
in \fref{fig:setup}. We assume that the atoms are trapped in
a 2D square lattice with exactly one atom per lattice site, as obtained
in a Mott insulator phase. The atoms are initially prepared in the
ground state, denoted by $\mid\downarrow\rangle$, and coherently
excited by a laser to a Rydberg state $\mid\uparrow\rangle$ with
Rabi frequency $\Omega$ and laser detuning $\Delta$ [see \fref{fig:setup}(a)].
Two atoms $i$ and $k$ both excited to the Rydberg state $\mid\uparrow\rangle$ and located at positions $\mathbf{r}_{i}$ and $\mathbf{r}_{k}$, respectively, interact via vdW interactions
$V(\mathbf{r}_{i}-\mathbf{r}_{k})=C_{6}(\theta_{i,k})/|\mathbf{r}_{i}-\mathbf{r}_{k}|^{6}$. These vdW interactions exceed typical ground state interactions of cold atoms by several orders of magnitude. We are interested in a situation where the vdW interaction has a non-trivial angular dependence $C_{6}(\theta_{i,k})$. Such an angular dependence arises, for example,
in laser excitation to higher angular momentum states, e.g. to Rydberg $p$-states, as opposed to excitation of $s$-states where the interactions are isotropic \cite{Reinhard2007}. In the remainder of this paper we will illustrate the anisotropic interactions by explicitly considering the stretched state $|n^2P_{3/2},m_j=3/2\rangle$ of Rubidium for which the $C_6(\theta_{i,k})$ is dominated by a term proportional to $\sin^4\theta_{i,k}$ \cite{Glaetzle2014}. Interactions are therefore much stronger along the $x$ direction compared to the $z$ direction [see \fref{fig:setup}(b)]. The atomic physics underlying
this interaction will be discussed in detail in \sref{sec:atomic} below. 

In its simplest form the dynamics of the driven Rydberg gas can be
described by an interacting system of pseudospin-1/2 particles

\begin{equation}
{H}=\frac{\hbar}{2}\sum_{i=1}^N\left(\Omega\sigma^{(i)}_{x}-\Delta\sigma^{(i)}_{z}\right)+\frac{1}{2}\sum_{i=1}^N\sum_{\substack{k=1,  k\neq i}}^N \frac{C_{6}(\theta_{i,k})P_{i}P_{k}}{\mid\mathbf{r}_{i}-\mathbf{r}_{k}\mid^{6}}
\label{eq:H}
\end{equation}
where $\sigma^{(i)}_{x}=\mid\uparrow\rangle_{i}\langle\downarrow\mid+\mid\downarrow\rangle_{i}\langle\uparrow\mid$
and $\sigma^{(i)}_{z}=\mid\uparrow\rangle_{i}\langle\uparrow\mid-\mid\downarrow\rangle_{i}\langle\downarrow\mid$
correspond to the local Pauli matrices and $P_{i}=\mid\uparrow\rangle_{i}\langle\uparrow\mid$
is the projection operator on the Rydberg level. We note that in this
model atoms are assumed to be pinned to the lattice sites, which is
referred to as the {\it frozen gas approximation} \cite{Low2012}. For isotropic
interactions spin models of this type have been discussed in previous
theoretical work \cite{Robicheaux2005,Weimer2008,Schachenmayer2010,Pohl2010,Lesanovsky2011,Garttner2012},
and have been the basis for interpreting experiments \cite{Schauss2012,Schauss2014}.

The modeling of the laser excited Rydberg gas as a coherent
spin dynamics governed by the Hamiltonian \eref{eq:H} is valid for
\emph{sufficiently short times}. First, as we noted above, the model
\eref{eq:H} ignores the motion of the atoms: laser excited Rydberg
atoms are typically not trapped by the optical lattice for the ground
state atoms, and there will be (large) mechanical forces associated
with the vdW interactions. In addition, Rydberg states have
a finite life time, scaling as $\tau\sim n^{3}$ $(\tau \sim n^5)$ for low (high) angular momentum states with $n$ the principal quantum number, and black body radiation can drive transitions between different Rydberg states, further decreasing the lifetime by approximately a factor of two \cite{Beterov2009}. The regime of validity has been analyzed in detail in \cite{Glaetzle2012}: there the long time dynamics of laser excited Rydberg gas including motion and dissipation was treated, including the validity of the frozen gas approximation and the cross over regime.

We emphasize that the various quantum phases predicted by the spin
model \eref{eq:H} as a function of the laser parameters and interactions,
and their preparation in an experiment, can only be understood in
a dynamical way. In an experiment all atoms are initially prepared
in their atomic ground state, \mbox{$|\Phi(t_0)\rangle=| G\rangle \equiv \mid\downarrow\rangle_1\ldots \mid\downarrow\rangle_N$}, which is the ground
state of the many-body Hamiltonian \eref{eq:H} for $\Omega=0$ and
$\Delta<0$. Preparation
of the ground state of the spin Hamiltonian \eref{eq:H} for a given
$\Omega$ and $\Delta$ can thus be understood in the sense of adiabatic
state preparation, where starting from an initial time $t_0$ with laser parameters $\Omega(t_{0})=0$ and $\Delta(t_{0})<0$
we follow the evolution of the many body state for a parameter trajectory
to the final time $t_f$ with $\Omega(t_{f})=\Omega$ and $\Delta(t_{f})=\Delta$, see \fref{fig:setup}(c).
This dynamical preparation of many-body states and quantum phases
of the spin-model \eref{eq:H} in a time-dependent mean field ansatz,
in particular in the anisotropic case, will be a central question
to be addressed below.

While our focus below will be on the anisotropic model, we find it
worthwhile to summarize the basic properties and signatures of the
quantum phases (ground states) of the spin model \eref{eq:H} for isotropic interactions. Even
for this case, the ground-state phase diagram
of the Hamiltonian \eref{eq:H} is rather complicated. In the so-called
\emph{classical limit}, $\Omega\rightarrow0$, where all terms in the Hamiltonian
\eref{eq:H} commute, the ground-state corresponds to the minimum
energy configuration of classical charges on a square lattice interacting
via a $1/r^{6}$ potential, and $\Delta$ serves as a chemical potential \cite{Rademaker2013}. As noted above,
for $\Delta<0$ this corresponds to the state $|G\rangle$ with all atoms in the ground state. For $\Delta>0$ a finite density of excited Rydberg atoms is
energetically favorable and the competition between the laser detuning
and the vdW interactions results in a complex crystalline arrangement
with a typical distance between excited atoms set by the length scale $\ell\equiv[C_{6}/(\hbar\Delta)]^{1/6}$. In two dimensions the Rydberg
atoms ideally want to form a triangular lattice to maximize their
distance, which will compete with the square optical lattice for the
setup of \fref{fig:setup}. An exact solution of this classical commensurability
problem is not known except in one dimension, where all possible commensurate
crystalline phases form a complete devil's staircase \cite{Bak1982}.
In analogy to the 1D case we expect a plethora of different crystalline
phases in two dimensions, which are stable over some part of the phase
diagram and which break the lattice symmetries in different ways \cite{Lee2001,Zeller2012,Rademaker2013}.

Away from the classical limit, $\Omega\neq0$, the crystalline states of
Rydberg atoms are expected to be stable for sufficiently small $\Omega$.
By increasing $\Omega$ quantum fluctuations will eventually melt
the crystalline phases and we reach a quantum disordered phase. The
nature of the corresponding quantum phase-transition has been studied
in one dimension \cite{Weimer2010,Sela2011} and remains an open issue
in higher dimensions.

Concerning anisotropic interactions, it is natural to expect that
the angular dependence of the vdW coefficient, $C_6(\theta_{i,k})$, is responsible for the
presence of an anisotropic crystalline phase at small $\Omega$. In
summary, our goal below is to describe the dynamical formation of
such crystals in large but finite systems similar to realistic experimental
situations, where finite size effects still play an important role,
and to compare the final state to the ground state of the system in
order to assess the fidelity of the dynamical preparation. To this
end, we developed an approach based on a time-dependent variational
principle which proved very useful to describe the crystalline states
for isotropic as well as anisotropic interactions with a large number
of excitations, i.e. in a parameter regime where an exact solution cannot be applied.

\subsection{Time dependent variational ansatz for many-body systems\label{sec:ansatz}}

In the following we present our variational ansatz and the corresponding equations of motion which we use to describe the dynamical preparation of Rydberg crystals governed by \eref{eq:H}. The simplest variational ansatz which is able to describe crystalline states of Rydberg atoms takes the most general product state form
\begin{eqnarray}
|\Phi(t)\rangle & = & \bigotimes_{i=1}^N\left[\alpha_{i}(t)\mid\downarrow\rangle_{i}+\beta_{i}(t)\mid\uparrow\rangle_{i}\right],
\label{eq:ansatz}
\end{eqnarray}
where $N$ denotes the number of atoms and the coefficients $\alpha_{i}$ and $\beta_{i}$ obey the normalization condition $|\alpha_{i}|^{2}+|\beta_{i}|^{2}=1$.
Crystalline phases correspond to states where the probability $|\beta_i|^2$ to find an atom in the Rydberg state at lattice site $i$ is spatially modulated and its Fourier components serve as an infinite set of order parameters. In contrast, in the quantum disordered phase the Rydberg density is homogeneous and $|\beta_i|^2$ is equal on all lattice sites $i$.

\subsubsection{Equilibrium properties of the variational ansatz\newline\newline}

Before deriving equations of motion for the time dependent variational parameters $\alpha_{i}(t)$ and $\beta_{i}(t)$ we discuss equilibrium properties of our variational ansatz.
Note that the ansatz \eref{eq:ansatz} captures the exact ground- and excited states of our model Hamiltonian [equation \eref{eq:H}] in the classical limit $\Omega \to 0$, where all eigenstates are product states. In the general case $\Omega>0$, it is an approximation and its validity will be discussed at the end of this subsection. In principle we find the variational ground-state by minimizing the expectation value of the Hamiltonian with respect to the variational parameters $\alpha_i$ and $\beta_i$. In the ground-state these parameters can be chosen to be real and the variational ground-state energy, $E=\langle\Phi|H|\Phi\rangle$, can be expressed as function of the probabilities $p_i=|\beta_i|^2$ as 
\begin{eqnarray}
E(p_{i}) &=& -\hbar\Omega\sum_{i}\sqrt{p_{i}(1-p_{i})} - \hbar\Delta \sum_i (p_{i}-\frac{1}{2}) \nonumber \\
&& + \frac{1}{2}\sum_{k\neq i} V(\mathbf{r}_{i}-\mathbf{r}_{k}) p_{i} p_{k} \label{eq:Evar}
\end{eqnarray}
with $V(\mathbf{r}_{i}-\mathbf{r}_{k})=C_{6}(\theta_{i,k})/|\mathbf{r}_{i}-\mathbf{r}_{k}|^{6}$. Finding all solutions of the corresponding mean-field equations $\partial E(p_{i})/\partial p_{i}=0$ in the thermodynamic limit is an impossible task, however. For this reason we do not attempt to make predictions about the phase diagram of \eref{eq:H} in the thermodynamic limit and rather focus on experimentally relevant systems with a finite but large number of atoms instead.

There is one notable exception, however: we can make a statement about the melting transition between the quantum disordered phase at large $\Omega$ and the adjacent crystalline phase within our variational (mean-field) approach. In the thermodynamic limit, the quantum disordered phase has a homogeneous Rydberg density $f_R := p_i \equiv p$ and we can determine at which point the homogeneous solution becomes unstable to density modulations.
Linearizing the mean-field equations in small perturbations around the homogeneous solution we find the condition 
\begin{equation}
1+\frac{\hbar^{2}\Omega^{2}}{\big[(V	_{0}f_{R}-\hbar\Delta)^{2}+\hbar^{2}\Omega^{2})\big]^{3/2}}\ \min_{\mathbf{k}}(V_{\mathbf{k}})=0 \ ,
\label{eq:boundary}
\end{equation}
where $V_{\mathbf{k}}=\sum_{i}e^{\iu\mathbf{k}\cdot\mathbf{R}_{i}}V(\mathbf{R}_{i})$
are Fourier components of the interaction potential (note that the
density $f_{R}$ of Rydberg excitations depends on $\Omega$ and $\Delta$).
We note that an expansion in small density modulations around the
homogeneous solution implicitly assumes that the melting transition
is continuous. It is possible that this transition could be first order,
however. In order to rule out a discontinuous melting transition we
minimized the variational energy \eref{eq:Evar} numerically on a lattice with $N=441$ sites
and found that the melting transition is indeed continuous.

The momentum $\mathbf{k}_0$ at which the interaction potential $V_{\mathbf{k}}$ is minimal determines the wave-vector at which density modulations form in the crystalline phase. In the isotropic case this minimum is at $\mathbf{k}_0=(k_0^x,k_0^z)=(\pi/a,\pi/a)$, where $a$ is the lattice
constant of the optical lattice. If one approaches the crystalline
phase from the quantum disordered phase, the leading instability is
thus always towards a crystalline state with Neel-type order, which
breaks a $Z_{2}$ lattice symmetry. Only at smaller $\Omega$ more
complicated crystalline states appear, which are most likely separated
by first order phase transitions. Consequently, within our variational approach the quantum phase
transition between the disordered and the crystalline phase is always
in the Ising universality class for isotropic interactions, independent of $\Omega$ and~$\Delta$.

For an anisotropic interaction potential with angular dependence $C_6( \theta_{i,k})$, present for example between $\mid \uparrow\rangle=|n^2P_{3/2},m_j=3/2\rangle$ states  of Rubidium as discussed in \sref{sec:atomic}, the minimum
is at a wave-vector $\mathbf{k}_0=(\pi/a,0)$.
Again, if we approach the crystalline phase from the quantum-disordered
regime, crystalline order will form only in $x$-direction with a period
of two lattice spacings, whereas no crystalline order is present in
$z$-direction. This transition is again continuous. The system thus
decouples into an array of quasi one dimensional Rydberg gases. Upon
further decreasing $\Omega$, we expect a transition to a state with
incommensurate, floating crystalline order in $z$-direction, in analogy
to one-dimensional systems \cite{Weimer2010,Sela2011}. The system remains long-range ordered
in $x$-direction, however, and finally settles into a commensurate,
genuinely two-dimensional crystalline state at sufficiently small
$\Omega$. We leave a detailed investigation of this two-step directional
melting transition open for future study.

\begin{figure}
\begin{centering}
\includegraphics[width=0.75\columnwidth]{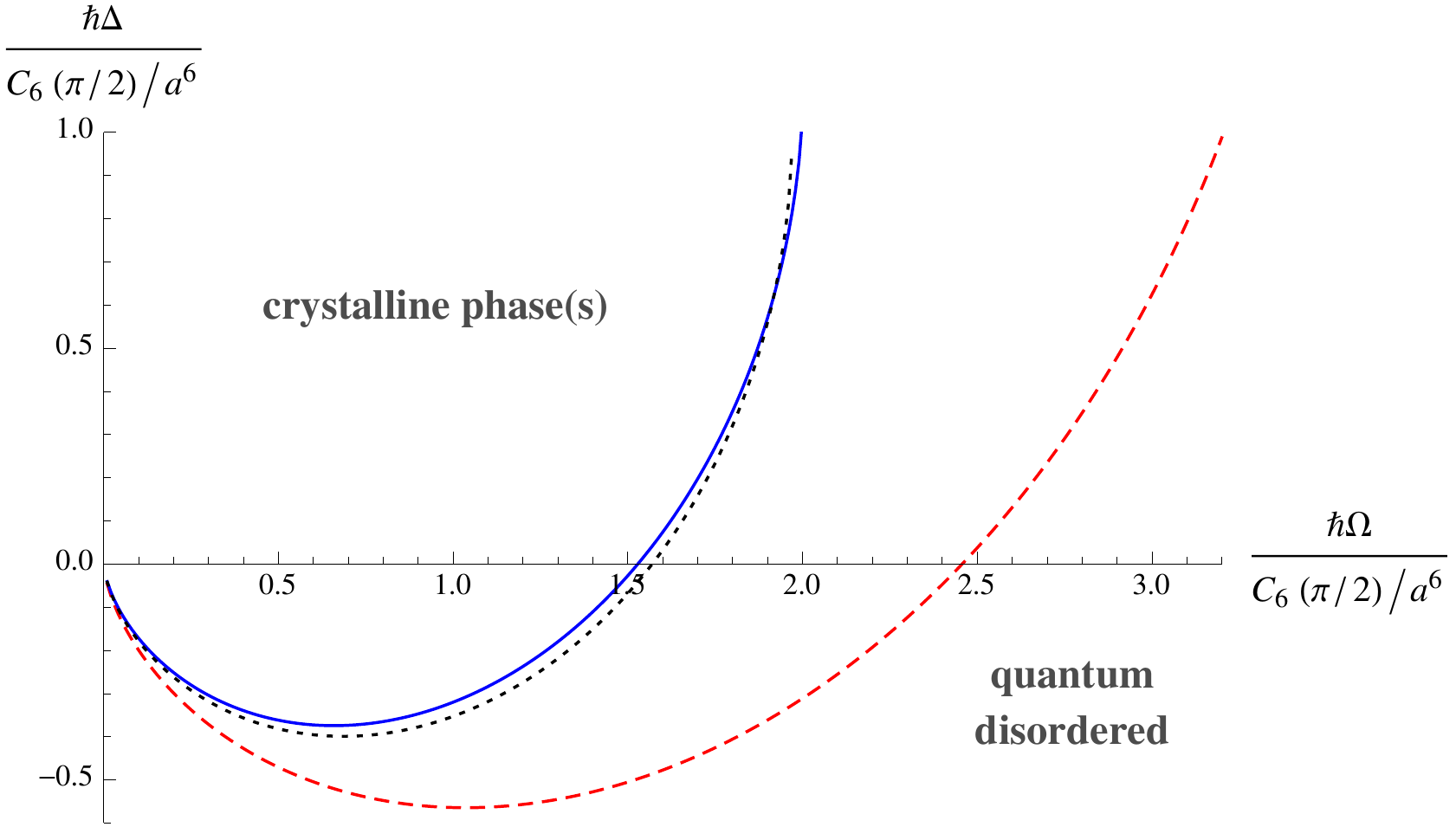} 
\protect\caption{Mean-field phase boundary between the quantum-disordered and crystalline phase(s) obtained from \eref{eq:boundary}. The blue solid line represents the case of anisotropic interactions with the angular dependence $C_6(\theta)$ of Rubidium $|nP_{3/2},m_j=3/2\rangle$ atoms [see \fref{fig:setup}(c)]. The red dashed line shows the phase boundary for isotropic interactions. For comparison, the black dotted line represents the mean-field phase boundary of a one-dimensional system. }
\label{fig:phase}
\end{centering}
\end{figure}

The phase boundary obtained from \eref{eq:boundary} is shown in \fref{fig:phase} for both isotropic as well as anisotropic interactions with the angular dependence represented in \fref{fig:setup}(b). As in the anisotropic case the interactions are much stronger in the $x$ direction compared to the $z$-direction, the corresponding phase-boundary significantly differs from the isotropic curve and is very close to the one obtained for a one-dimensional system.
Note that the mean-field phase boundary has an unphysical re-entrance behavior at negative detunings. This is because our variational ansatz vastly overestimates the ground-state energy in the quantum disordered phase at finite $\Omega$, where pair-correlations are important. 

Indeed, our product ansatz of \eref{eq:ansatz} does not
describe correlations between local density fluctuations of Rydberg atoms 
\begin{equation}
\langle\delta P_{i}\delta P_{k}\rangle=\langle P_{i}P_{k}\rangle-\langle P_{i}\rangle\langle P_{k}\rangle\ ,
\end{equation}
and thus $\langle\delta P_{i}\delta P_{k}\rangle_{\Phi}\equiv0$ for
our variational wave-function. Deep in the crystalline phase these
correlations are weak and decay exponentially with distance, however.
We can give an upper bound on the strength of such density density
correlations and consequently make a statement about the validity
of our ansatz by estimating the strength of local density fluctuations
\begin{equation}
\langle{\delta P_{i}}^{2}\rangle=\langle P_{i}\rangle(1-\langle P_{i}\rangle)\ .
\end{equation}
Accordingly, density-density correlations are negligible deep in the
crystal, where $\langle P_{i}\rangle$ is either close to zero or
one. As a consequence we expect our ansatz to be valid also at finite
$\Omega$ as long as we are deep in the crystalline phase. At sufficiently large $\Omega$, where the system enters the
quantum disordered phase and quantum correlations become predominant, 
our ansatz is not a good approximation for the exact ground-state wave function. 
 
\subsubsection{Time-dependent variational ansatz and Euler-Lagrange equations\newline\newline}

One of the main goals of our paper is to describe the dynamical formation of crystalline states of Rydberg excitations in a large but finite system during a slow change of the laser parameters. For this reason we incorporate our ansatz into a time-dependent variational approach,
where the formation of Rydberg excitations is described by the time evolution of the variational coefficients $\alpha_i(t)$ and $\beta_i(t)$ and governed by the Hamiltonian \eref{eq:H}. Considering an initial condition where all atoms are in the ground state $|G\rangle$, i.e. $\alpha_i(t_0)=1\ (\forall i \in \{1,\dots,N\})$ and $\Omega(t_0)=0$, $\Delta(t_0)<0$, we use the time-dependent variational principle (TDVP) \cite{Kramer2008} to derive the equations of motion for the variational coefficients during a slow change of $\Omega$
and $\Delta$ as in typical dynamical state preparation schemes \cite{Pohl2010,Schachenmayer2010,Schauss2014}.
The TDVP states that the time-evolution of the variational coefficients satisfy the least action principle which means that they can be derived using the Euler-Lagrange equations:
\begin{eqnarray*}
\frac{d}{dt}\left(\frac{\partial L}{\partial\dot{\alpha}_{i}^{*}}\right) & = & \frac{\partial L}{\partial\alpha_{i}^{*}}\\
\frac{d}{dt}\left(\frac{\partial L}{\partial\dot{\beta}_{i}^{*}}\right) & = & \frac{\partial L}{\partial\beta_{i}^{*}}
\end{eqnarray*}
where  $L$ is the Lagrangian 
\begin{eqnarray*}
L & = & \frac{\iu\hbar}{2}\langle\Phi|d_{t}\Phi\rangle-\frac{\iu\hbar}{2}\langle d_{t}\Phi|\Phi\rangle-\langle\Phi|H|\Phi\rangle.
\end{eqnarray*}
leading to a set of $2N$ non-linear coupled equations
\begin{eqnarray}
\iu\hbar\dot{\alpha_{i}} & = & \frac{\hbar\Omega}{2}\beta_{i}+\frac{\hbar\Delta}{2}\alpha_{i}\nonumber \\
\iu\hbar\dot{\beta_{i}} & = & \frac{\hbar\Omega}{2}\alpha_{i}-\frac{\hbar\Delta}{2}\beta_{i}+\sum_{k\neq i}\frac{C_{6}(\theta_{i,k})}{|\mathbf{r}_{i}-\mathbf{r}_{k}|^{6}}|\beta_{k}|^{2}\beta_{i}.\label{eq:eq_motion}
\end{eqnarray}
We note that these equations conserve the norm of the wavefunction for all times, i.e. $|\alpha_{i}(t)|^{2}+|\beta_{i}(t)|^{2}=1$. If $\Omega$ and $\Delta$ do not evolve in time, the expectation value of the energy $E=\langle\Phi|H|\Phi\rangle$ is also a conserved quantity. However, this is not the case for a \emph{dynamical} state preparation and the final energy depends crucially on the parameter trajectory.

In a perfectly adiabatic situation we would obtain the variational ground-state for
$\Omega(t_f)$ and $\Delta(t_f)$ at the end of the time evolution.
Since our sweep protocols are limited to timescales smaller than the
lifetime of the Rydberg state, the preparation will not be perfectly
adiabatic and we discuss deviations from adiabaticity in sub\sref{sec:testcase}. We note that given the phase diagram shown in \fref{fig:phase} and the typical parameter sweep we consider [\fref{fig:setup}(b)], the system has to undergo a quantum phase transition from the quantum disordered phase to the crystalline phase at some point in the preparation which may reduce the adiabaticity of the preparation significantly. In the finite systems that we consider in the rest of this work, a finite-size gap is always present which reduces this problem, however.

Finally, we note that our ansatz \eref{eq:ansatz} is particularly
suited to study the experimentally relevant situation where the Rydberg laser is switched off at the end of the parameter sweep $\Omega(t_f)=0$, because it captures the exact ground- and excited states of the Hamiltonian \eref{eq:H} in the classical limit $\Omega=0$, as discussed above.  In sub\sref{sec:testcase} we also estimate the typical Rabi frequency $\Omega$ at which our ansatz fails by comparing our approach to the exact solution of the Schrödinger equation.

In the next section, we explain in detail the implementation of the model Hamiltonian \eref{eq:H} with Rydberg atoms excited to $\mid\uparrow\rangle=|n^2P_{3/2},m_j=3/2\rangle$ Rydberg states in order to provide realistic parameters for our numerical \sref{sec:results}.

\section{Anisotropic interactions for Rydberg atoms in p-states\label{sec:atomic}}
In the following we discuss the derivation of our model Hamiltonian [equation \eref{eq:H}] from a microscopic Hamiltonian,
\begin{equation}
H_{\rm mic}=\sum_{i=1}^N\left[H_A^{(i)}+H_L^{(i)}\right]+\frac12 \sum_{i=1}^N\sum_{\substack{k=1 , k\neq i}}^N H_V^{(i,k)}.
\label{eq:micro}
\end{equation}
describing vdW interactions between $N$ alkali atoms laser excited to the \mbox{$\mid \uparrow\rangle\equiv |nP_{3/2},m_j=3/2\rangle$} Rydberg state. 
We first focus on the Rydberg manifold and their interactions and then discuss the laser excitations. The first term of \eref{eq:micro},
\begin{equation}
H_A^{(i)}=\sum_{m_j}\left[\hbar \omega_{nP_{3/2}}+\mu_B g_j B_zm_j \right] |m_j\rangle\langle m_j|,
\end{equation}
accounts for the energies of the Zeeman sublevels \mbox{$|m_j\rangle\equiv |nP_{3/2},m_j\rangle$} with \mbox{$m_j\in\{-3/2,\ldots,3/2\}$} as illustrated in \fref{fig:setup}. Here, $\hbar \omega_{nP_{3/2}}$ is the energy difference between the atomic ground state, $5S_{1/2}$, and the $nP_{3/2}$ Rydberg manifold in the absence of external fields. The second term of $H_A^{(i)}$ describes the lifting of the energy degeneracy of the $nP_{3/2}$ Rydberg manifold due to a magnetic field $\mathbf{B}=B\mathbf{e}_z$ (see \fref{fig:setup}), with $\mu_B/h= 1.4$ MHz/G the Bohr magneton and $g_{j}$ the Lande factor for $j=3/2$. Note, that the quantization axis of the corresponding eigenstates is given by the direction of the magnetic field, $\mathbf{B}$, and is aligned {\it in plane} along the $z$-axis, see \fref{fig:setup}. In order to neglect higher order shifts and to prevent mixing between different fine-structure manifolds the energy shifts $\Delta E_{m_j}=\mu_B g_j B_zm_j  $ have to be much smaller than the fine-structure splitting $E_{nP_{3/2}}-E_{nP_{1/2}}$ of the Rydberg manifolds. Typically, the fine structure splitting is of the order of tens of GHz, e.g. 7.9 GHz for $n=25$.

Away from Foerster resonances two laser excited Rydberg atoms dominantly interact via van der Waals interactions \cite{Saffman2010,Low2012}. In general, these van der Waals interactions, $\hat V_{\rm vdW}$, will mix different Zeeman sublevels $|m_j\rangle$ in the $nP_{3/2}$ manifold \cite{Walker2008}. Let us denote by \mbox{$\hat P=\sum_{i,j}| m_i,m_j\rangle\langle m_i,m_j|$} a projection operator into the $nP_{3/2}$ manifold, then the  dipole-dipole interactions $\hat V_{\rm dd}$ will couple to intermediate states, \mbox{$\hat Q_{\alpha,\beta}= |\alpha,\beta\rangle\langle \alpha,\beta|$}, which have an energy difference $\delta_{\alpha\beta}$. In second order perturbation this gives rise to
\begin{equation}
\hat V_{\rm vdW}=\hat P\sum_{\alpha\beta}\frac{\hat V_{\rm dd}\hat Q_{\alpha,\beta}\hat V_{\rm dd}}{\delta_{\alpha\beta}}\hat P,
\label{eq:vdwop}
\end{equation}
where $\hat V_{\rm vdW}$ is understood as an operator acting in the manifold of Zeeman sublevels  We note, that in the absence of an external magnetic field, $\mathbf{B}\rightarrow 0$, the new eigenenergies obtained from diagonalizing $\hat V_{\rm vdW}$ are isotropic. 

Anisotropic van der Waals interactions can be obtained by lifting the degeneracy between the Zeeman sublevels e.g. with a magnetic field. For distances large enough, such that the off-diagonal vdW coupling matrix elements of \eref{eq:vdwop} are much smaller than the energy splitting between the Zeeman sublevels, it is possible to simply consider interactions between $|nP_{3/2}, m_j=3/2\rangle$ states and neglect transitions to different $m_j$ levels. Typically, for Rydberg $p$-states the off-diagonal vdW matrix elements are of the same order as the diagonal interaction matrix elements. Pairwise interactions between two atoms excited to the \mbox{$\mid\uparrow\rangle=|nP_{3/3},m_j=3/2\rangle$} state are then described by the Hamiltonian 
\begin{equation}
H^{(i,k)}_V=V(\mathbf{r}_i-\mathbf{r}_k)\mid\uparrow\rangle_i\langle \uparrow\mid\otimes  \mid\uparrow\rangle_k\langle \uparrow \mid,
\end{equation}
where $V(\mathbf{r}_i-\mathbf{r}_k)=\langle {\textstyle \frac32,\frac32 }|\hat V_{\rm vdW}|{\textstyle \frac32,\frac32 }\rangle=C_6(\theta_{i,k})/|\mathbf{r}_i-\mathbf{r}_k |^6$ the van der Waals interaction potential, giving rise to the second term of \eref{eq:H}. Here, $\theta_{i,k}=\sphericalangle (\mathbf{B},\mathbf{r}_i- \mathbf{r}_k)$ is the angle between the relative vector and the quantization axis given by the magnetic field direction $\mathbf{B}$ (see \fref{fig:setup}).  

The second term of \eref{eq:micro}, $H_L^{(i)}$, accounts for the laser excitation of Rubidium $^{87}$Rb atoms prepared in their electronic ground state, which we choose as \mbox{$\mid \downarrow \rangle=|5^2S_{1/2},F=2,m_F=2\rangle$}, to the \mbox{$\mid\uparrow\rangle=|nP_{3/2},m_j=3/2\rangle$} Rydberg states. This can be done using a single-photon transition with Rabi frequency $\Omega$, scaling as $\Omega\sim n^{-3/2}$. Using a UV laser source it is possible to obtain Rabi frequencies of several MHz in order to excite Rydberg states around $n\sim 30$. The single particle Hamiltonian governing the laser excitation of atom $i$ is 
\begin{equation}
H_L^{(i)}(t)=\frac{\hbar\Omega}{2}\left(\mid\uparrow\rangle_i\langle\downarrow\mid e^{i(\mathbf{k}_L\cdot \mathbf{r_i}-\omega_L t)}+\mid\downarrow\rangle_i\langle\uparrow\mid e^{-i(\mathbf{k}_L\cdot \mathbf{r_i}-\omega_L t)}\right),
\end{equation}
where $\omega_L=\omega_{nP_{3/2}}+\Delta E_{3/2}/\hbar+\Delta$ is the laser frequency detuned by $\Delta$ from the atomic transition (including the magnetic field splitting) and $\mathbf{k}_L$ is the wave vector of the laser. Unwanted couplings to Zeeman levels with $m_j\neq 3/2$ due to the laser can be prevented by using a detuning  $|\Delta|\ll |\Delta E_{3/2}-\Delta E_{1/2}|$ or by using circular polarized light propagating along the quantization axis, i.e. $\mathbf{k_L}\sim \mathbf{\hat z}$, which couples $\mid\downarrow\rangle$ only to the $m_j=3/2$ state. In a frame rotating with the laser frequency and after absorbing the position dependent phase into \mbox{$\mid\uparrow\rangle_i\rightarrow e^{-i\mathbf{k}_L\cdot \mathbf{r_i}}\mid\uparrow\rangle_i$} one obtains the first term of \eref{eq:H}.

As an example, we consider the $n=25$ Rydberg state, i.e.  \mbox{$\mid\uparrow\rangle\equiv |25P_{3/2},m_{j}=\frac{3}{2}\rangle$}. For the  corresponding diagonal van der Waals matrix element we obtain using the model potential from \cite{Marinescu1994}

\begin{equation}
C_{6}(\theta)=\left(6.33\sin^{4}\theta-0.267\sin^{2}\theta+0.269\right)h\, \mathrm{MHz}\,\mu \mathrm{m}^6.\label{eq:C6_n25}
\end{equation}
Thus, $C_{6}(\pi/2)=6.35\,h\, \mathrm{MHz}\,\mu \mathrm{m}^6$ and $C_{6}(0)=0.269\,h\, \mathrm{MHz}\,\mu \mathrm{m}^6$. The dominant $\sim \sin^4\theta$ term arises from dipole-dipole transitions to $nS_{1/2}$ states, while residual interactions at $\theta=0$ and deviations from the $\sim \sin^4\theta$ shape originate from couplings to $D$-channels as discussed in \cite{Reinhard2007}. The lifetime of the Rydberg state is $\tau\approx29\ \mu$s \cite{Beterov2009}.

We finally note that it is also possible to switch from the anisotropic configuration defined above to an isotropic configuration where the angle $\theta$ is fixed to a constant value $\frac{\pi}{2}$. In this configuration, the magnetic field is rotated from the $z$ to the $y$-direction (see \fref{fig:setup}) and the Rydberg state which is excited by the laser $\mid\uparrow\rangle$ is in this case $|n^2P_{3/2},m_j=\frac{3}{2}\rangle_y$.

\section{Numerical results\label{sec:results}}

In the following we present our numerical results obtained by propagating the equations of motion \eref{eq:eq_motion} along different parameter trajectories $(\Omega(t),\Delta(t))$ in order to describe the dynamical state preparation of Rydberg crystalline phases. To this end we first estimate the domain of validity of our approach based on the TDVP comparing in the case of small systems our results to the exact diagonalization (ED) solution which is obtained from the Schrödinger equation. We then present the results of our approach for large systems ($N>500$) with large densities of Rydberg excitations where ED techniques cannot be applied.

\subsection{Validity of the variational approach for small systems \label{sec:testcase}}

\begin{figure}
\begin{centering}
\includegraphics[width=0.75\columnwidth]{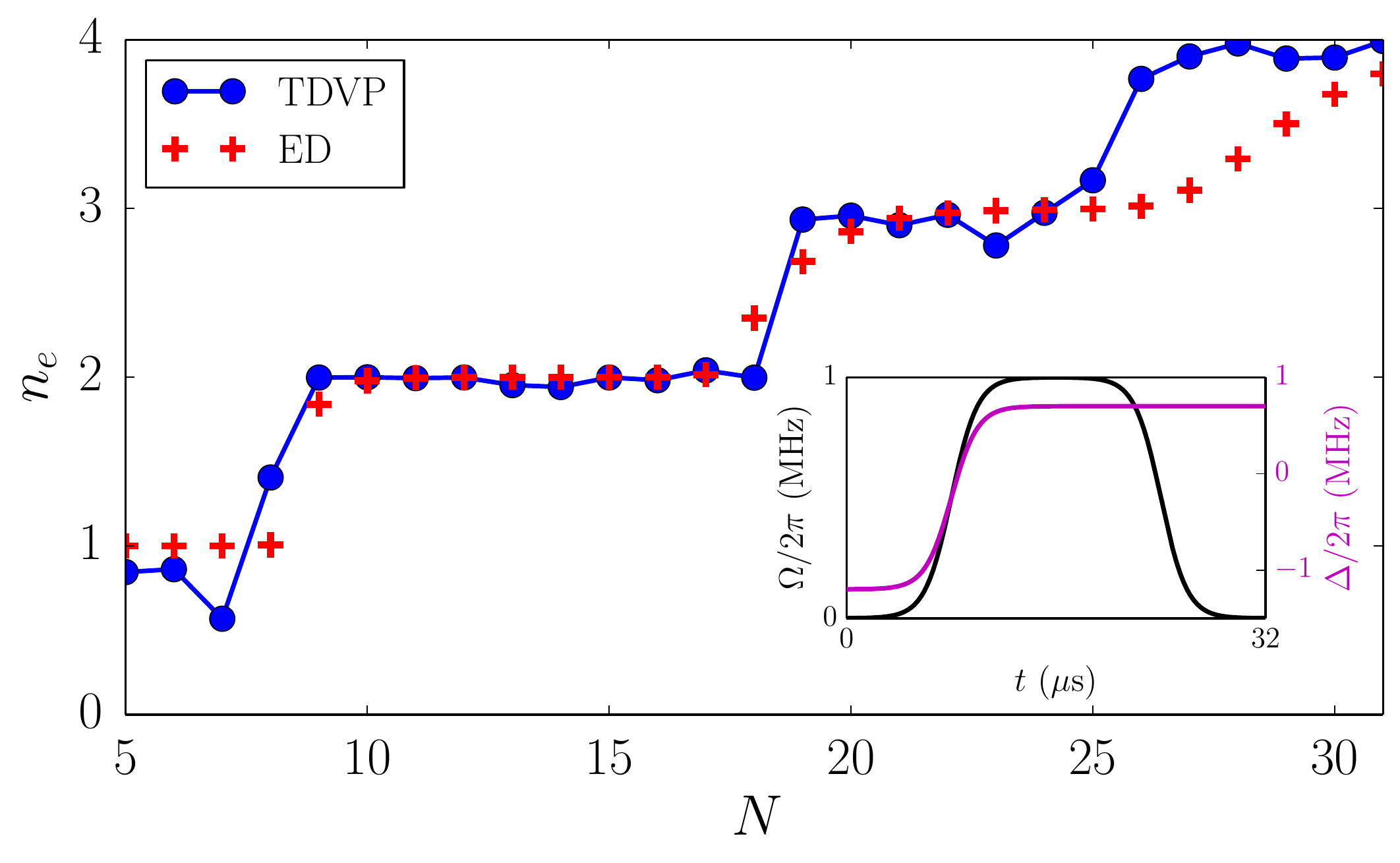}
\protect\caption{Number of excitations $n_e$ of the final state of the preparation as a function of the number $N$ of atoms obtained after a
sweep time of $32\ \mu s$ with $\Omega(t_f)=0$, $\Delta(t_f)/(2\pi)=0.7$ MHz and a Rydberg state $|43S_{1/2},m_{j}=\frac{1}{2}\rangle$.
Blue circles represent our variational  approach, red crosses the exact solution
obtained by the Schrödinger equation and the same sweep. Inset: Laser parameters $\Omega/(2\pi)$ and $\Delta/(2\pi)$ as a function of time $t$ used for the dynamical state preparation. \label{fig:staircase}}
\end{centering}
\end{figure}

It is instructive to start by considering small systems where an exact numerical solution is available which allows us to estimate the validity of our approach.
The first situation we have in mind is the classical limit ($\Omega=0$) of the model \eref{eq:H} for a one-dimensional system where the number of excitations takes the form of the stair case as a function of the number of atoms $N$ \cite{Bak1982}. As its existence was recently demonstrated experimentally \cite{Schauss2014} we consider as a first illustration of our approach the same parameters as in \cite{Schauss2014} with the notable exception that we choose a  larger sweep time $t_{f}=32\ \mu$s instead of $t_f=4\ \mu$s in order to describe the dynamical preparation of states which are as close as possible to the ground state of the system. We now test our approach by comparing the exact number of excitations $n_e$ for theses parameters, obtained after a truncation of the Hilbert space \cite{Pohl2010}  , to the one corresponding to our variational ansatz. The result is shown in \fref{fig:staircase} as a function of the number of atoms $N$ where the laser sweep is represented in the inset. Our approach describes very well the excitation stair case and apart from some defects (for example for $N=8$) compares very well with the exact solution (red crosses). We finally emphasize that as the sweep time is
increased, our solution converges towards the exact classical ground state, as expected. 

Our approach describes the key feature of the one-dimensional system but as it relies on a mean-field approximation of the many-body Hamiltonian \eref{eq:H}, its domain of validity may strongly depend on the dimension of the system. As a second illustration of our variational approach we consider, therefore, a small two-dimensional system of $N=16$ atoms, in an isotropic configuration where an ED solution based on the truncation of the Hilbert space is still available.

\begin{figure}
\begin{centering}
\includegraphics[width=0.75\columnwidth]{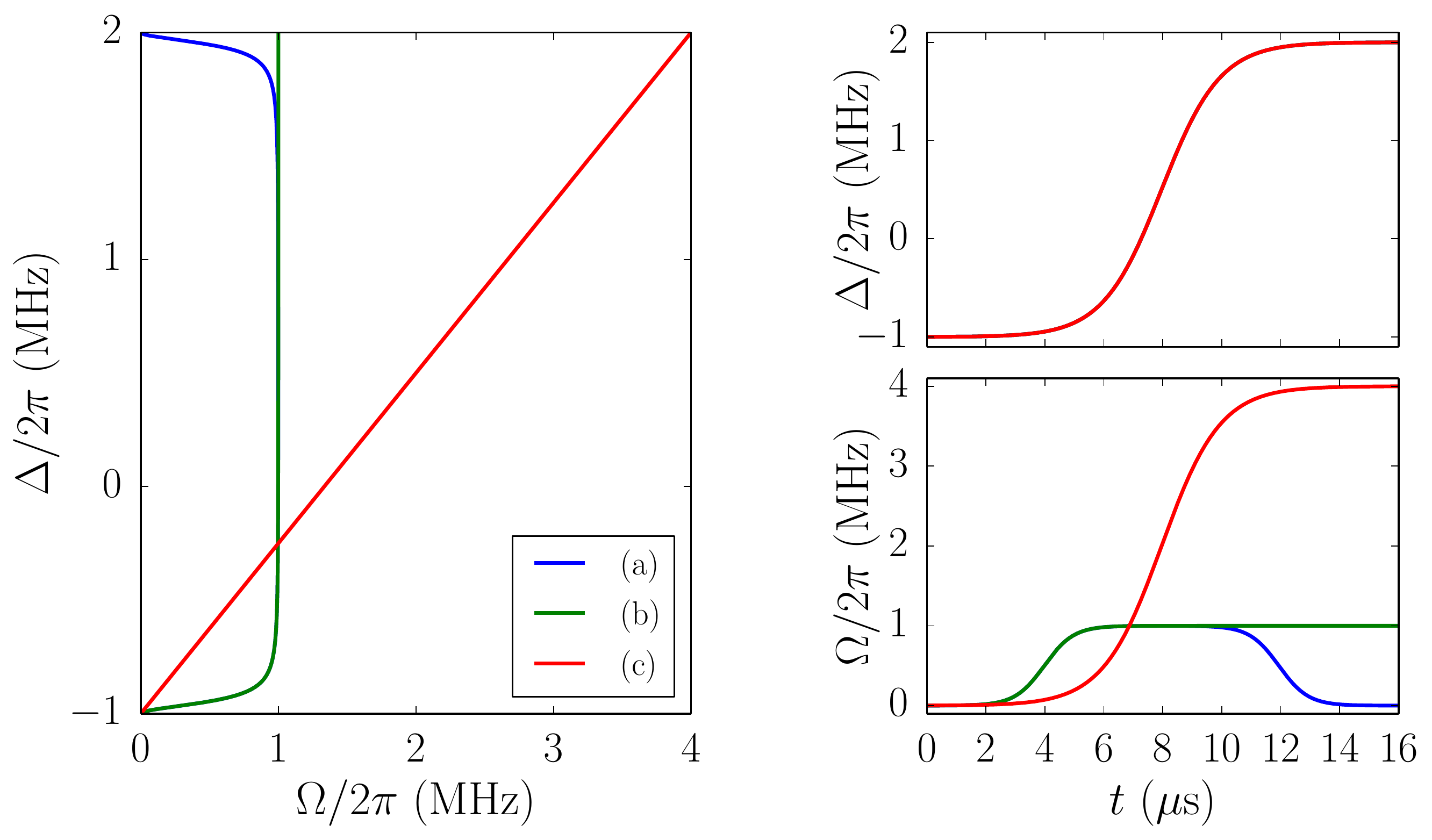} 
\protect\caption{The three different parameter sweep paths (a), (b), (c), corresponding
to the final Rabi frequencies $\Omega(t_f)/2\pi=0,1$ and $4$ MHz. The
left plot shows the three paths in parameter space, whereas the right
plots show how the parameters evolve as a function of time. Initially,
all atoms are in the ground state with $\Omega(t_0)=0$ and $\Delta(t_0)/(2\pi)=-1$
MHz. \label{fig:sketch}}
\end{centering}
\end{figure}

Our goal here is to show the influence of the final Rabi frequency $\Omega(t_f)$ on the validity of our approach. To this end, we consider three different sweeps of the Rabi-frequency $\Omega$ and detuning $\Delta$ along the paths shown in \fref{fig:sketch} corresponding to three final Rabi frequencies $\Omega(t_f)/(2\pi)=0,1,4$ MHz at a positive detuning $\Delta(t_f)/(2\pi)=2$ MHz. In all three cases we start at a negative detuning $\Delta(t_0)/(2\pi)=-1$ MHz and zero Rabi frequency and compute the mean distribution of excited
Rydberg atoms at the end of the sweep, which is given by $|\beta_{i}|^{2}$. We choose a lattice spacing $a=532$ nm and the Rydberg level $\mid\uparrow\rangle=|25P_{3/2},m_{j}=3/2\rangle$
corresponding to a $C_{6}$ coefficient \eref{eq:C6_n25} where due to the isotropic configuration considered here, the angle $\theta$ is fixed to $\frac{\pi}{2}$. The sweep time is $t_{f}=16\ \mu s$
which is lower than the lifetime of the Rydberg excitations $\tau\approx29\ \mu$s \cite{Beterov2009}. 

\begin{figure}
\begin{centering}
\includegraphics[height=3cm]{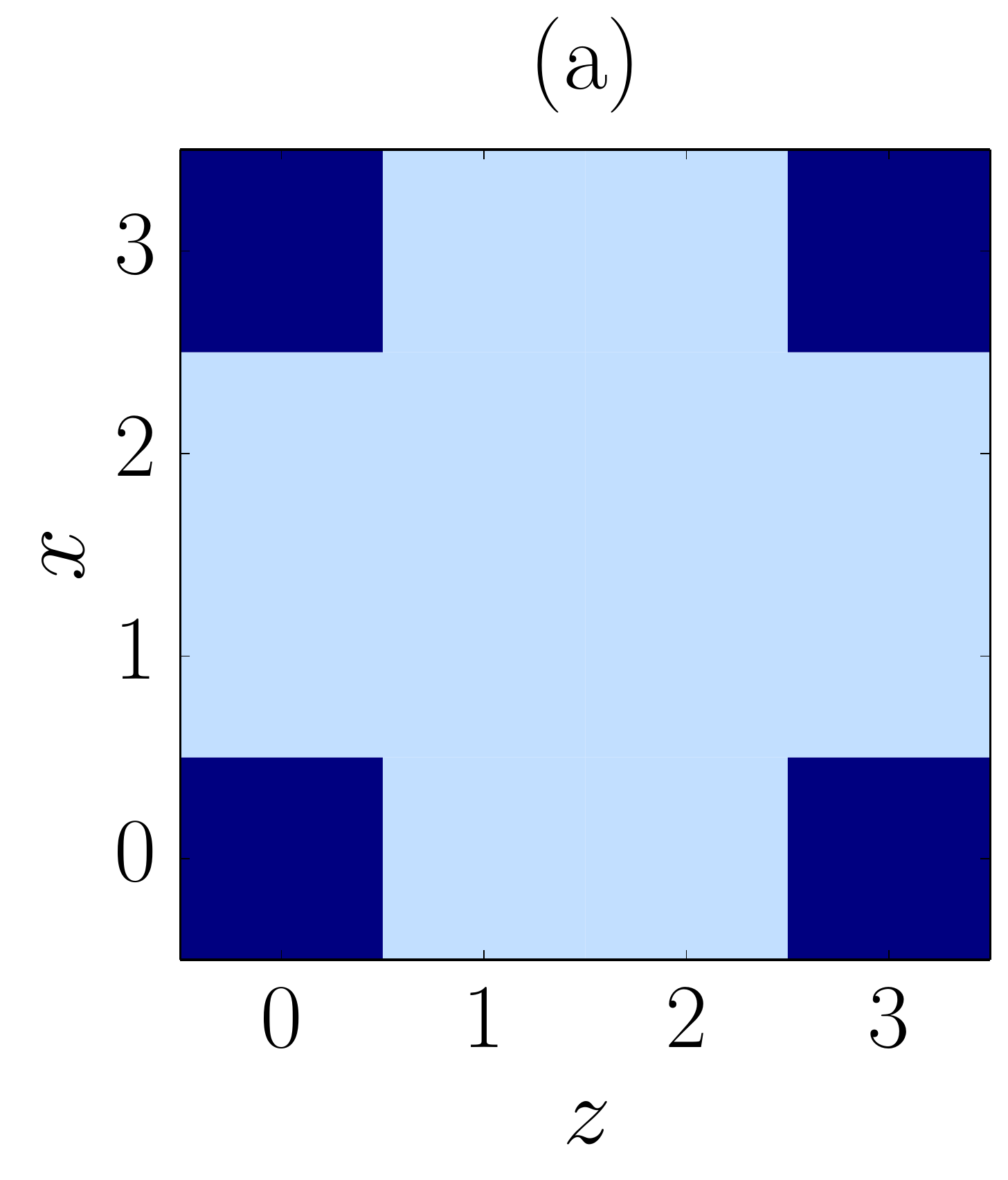}
\includegraphics[height=3cm]{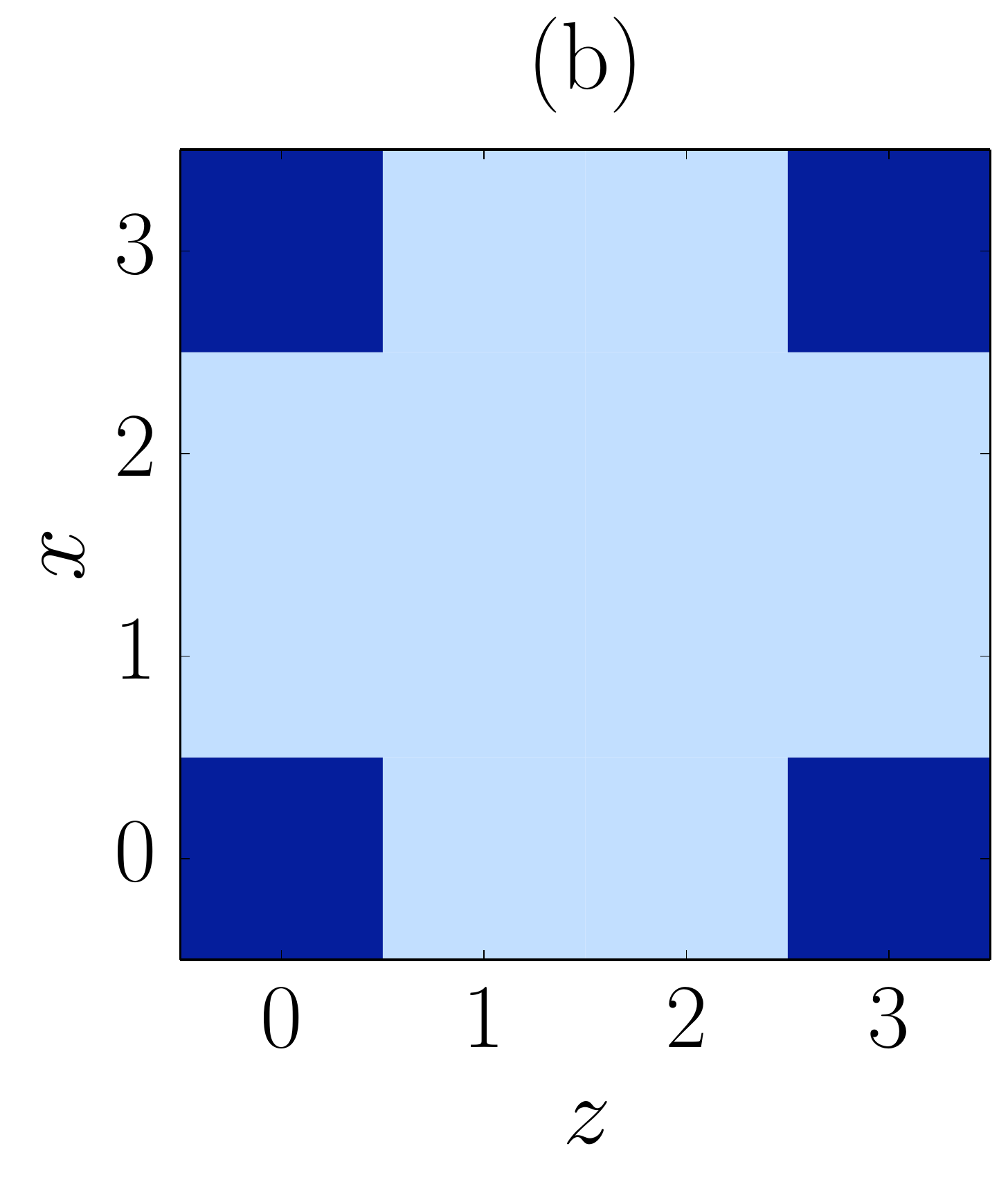}
\includegraphics[height=3cm]{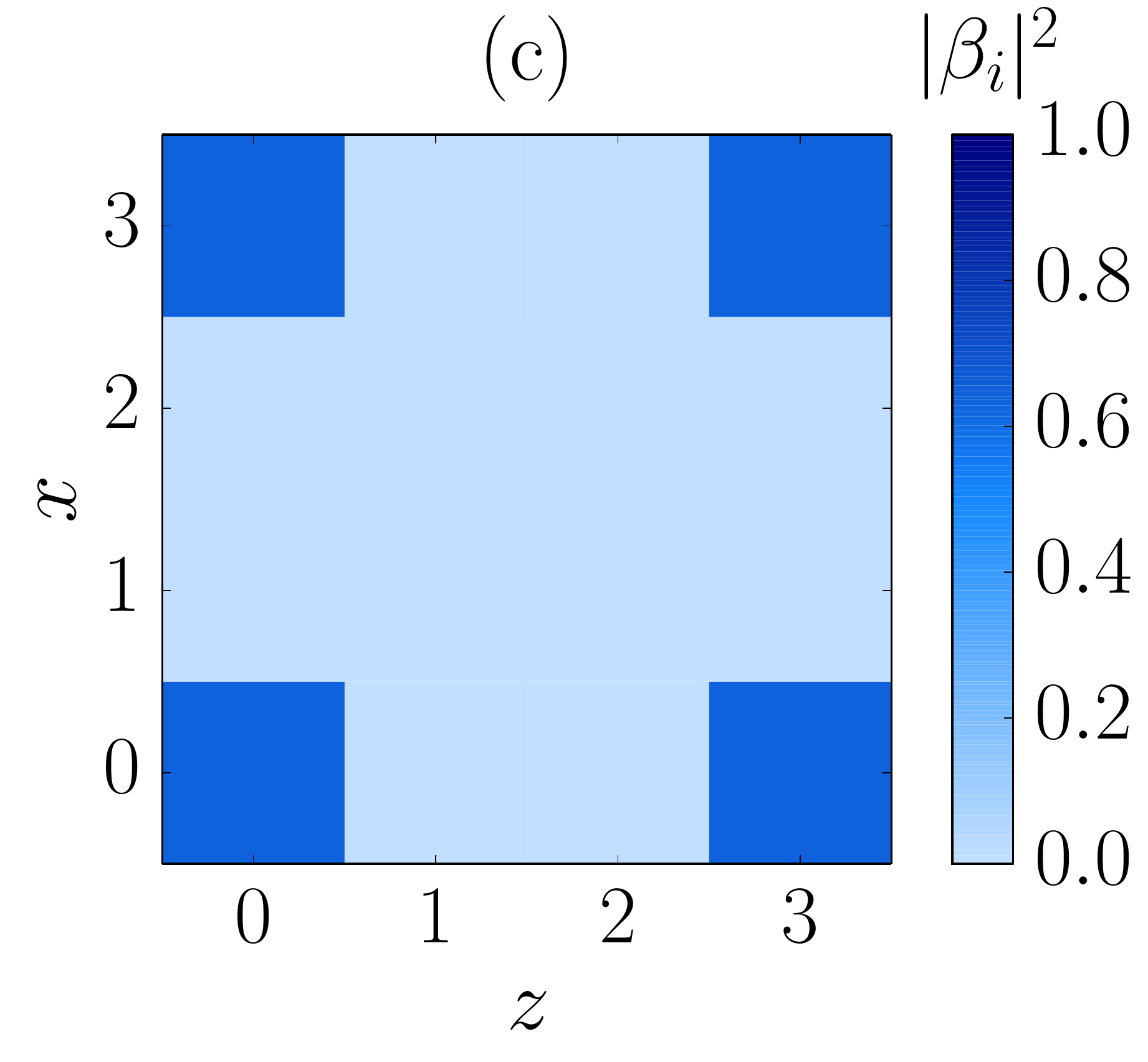}

\includegraphics[height=3cm]{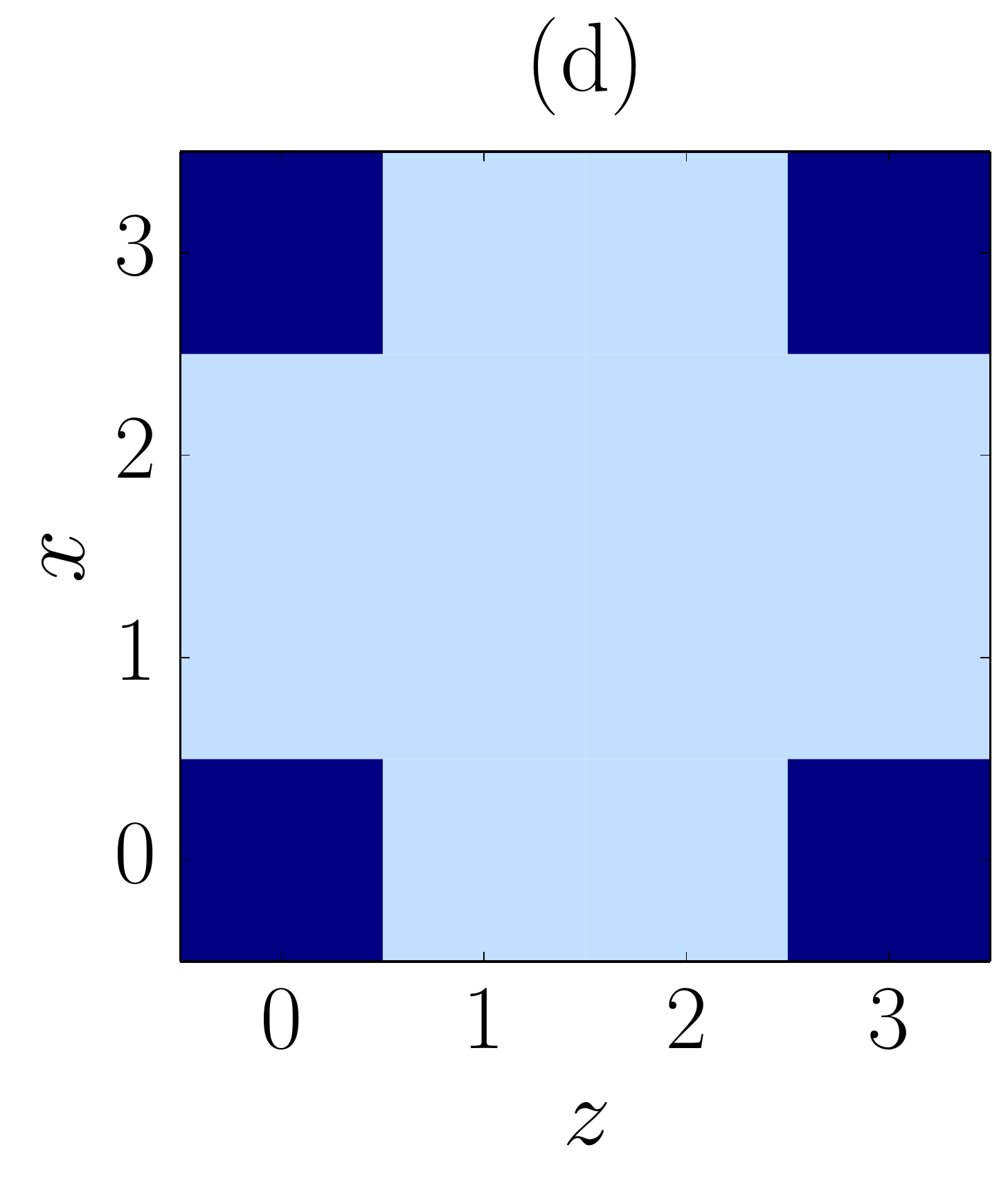}
\includegraphics[height=3cm]{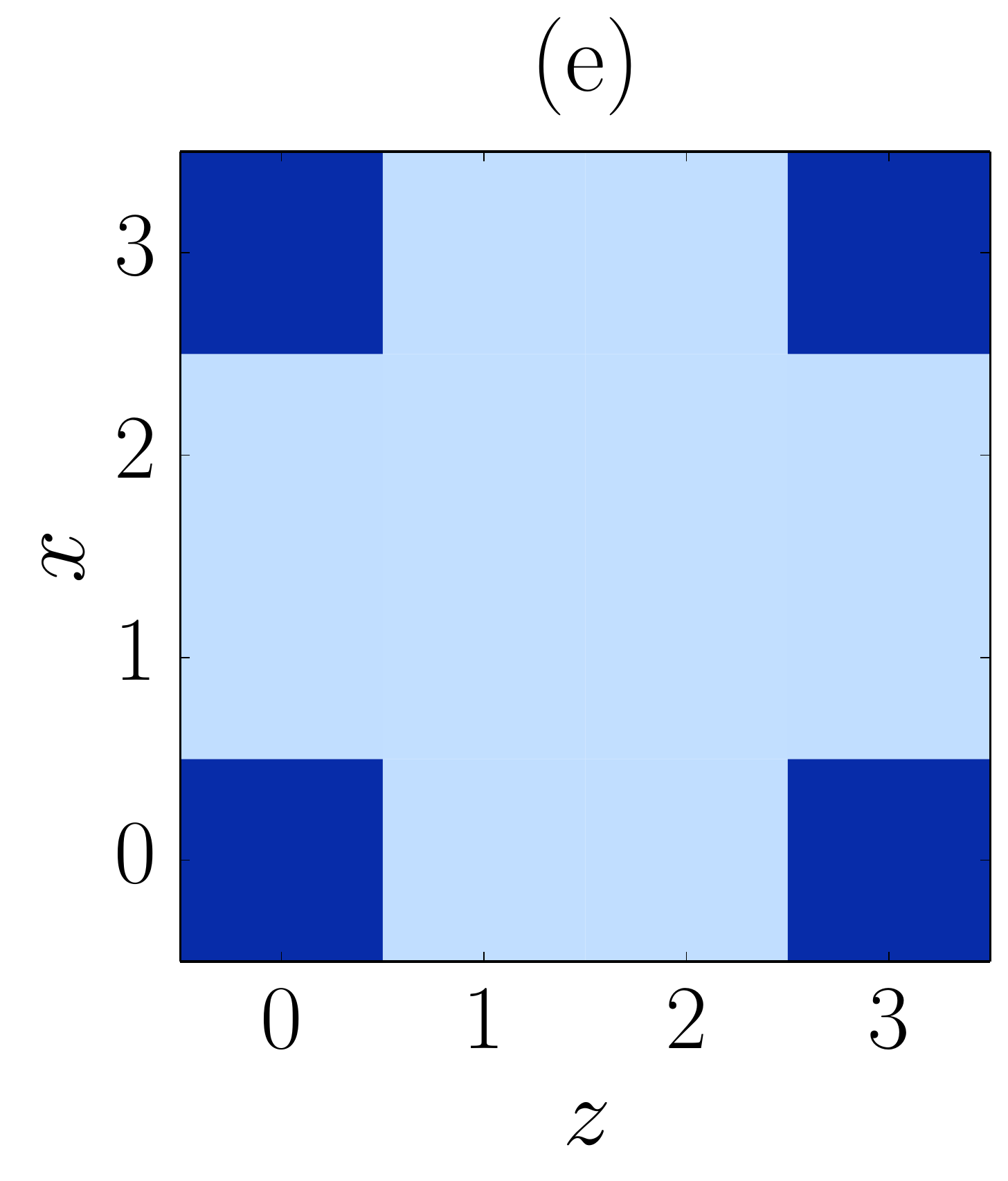}
\includegraphics[height=3cm]{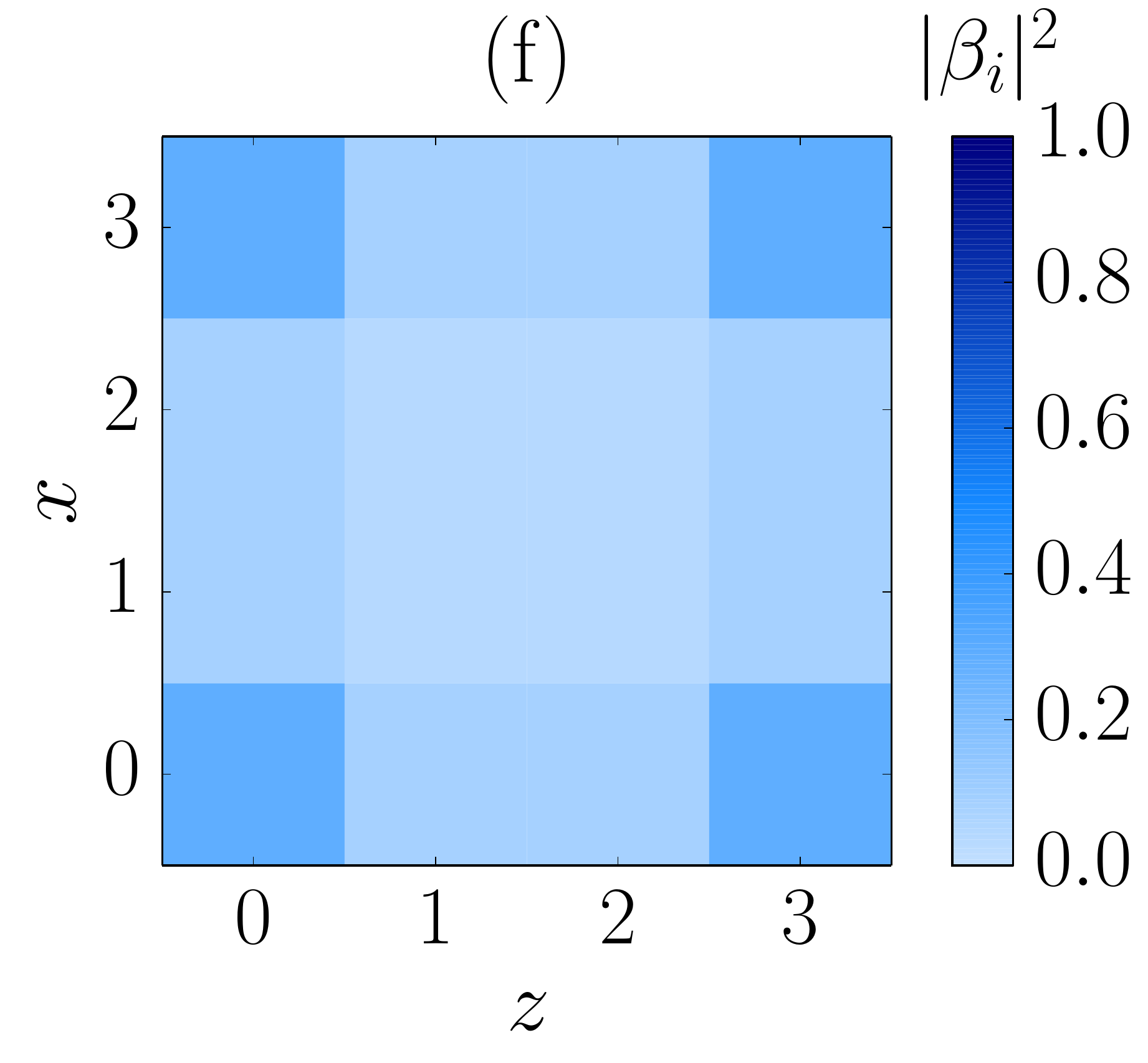} 
\protect\caption{Comparison of TDVP with ED. Upper panel: distribution
of Rydberg atoms for isotropic interactions after the three parameter
sweeps shown in \fref{fig:sketch} using the TDVP approach. Lower
panel: same as in the upper panel but calculated using ED.
For small final Rabi frequencies the results obtained by both approaches
are basically indistinguishable. The corresponding energies per particle
are $E=0.603,0.556,0.204\,h\,$MHz for (a), (b), (c) in the upper panel
and $E=0.602,0.548,-0.049\,h\,$MHz for (d), (e), (f) in the lower panel.
\label{comparison}}
\end{centering}
\end{figure}

\begin{figure}
\begin{centering}
\includegraphics[height=4cm]{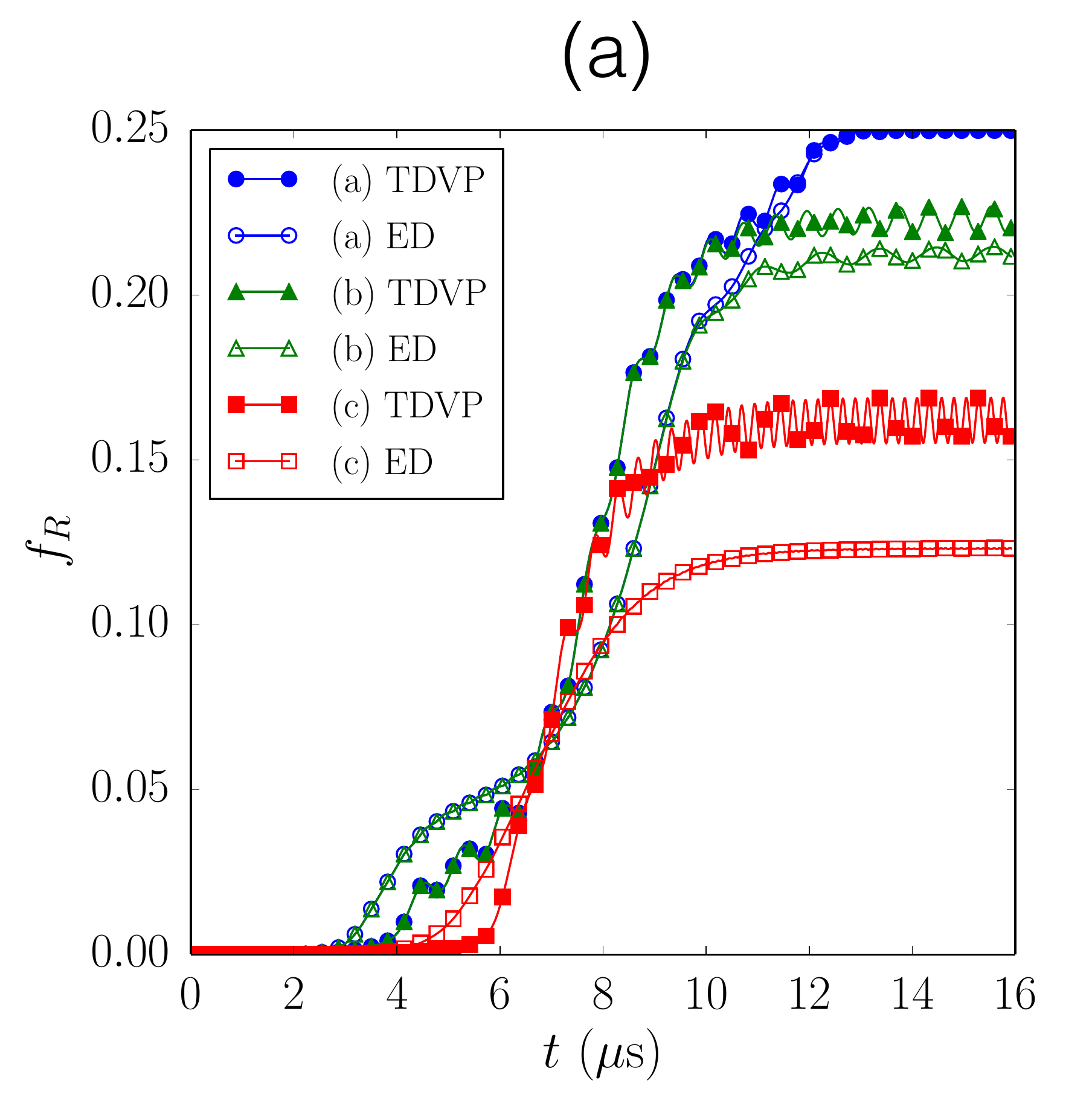}
\includegraphics[height=4cm]{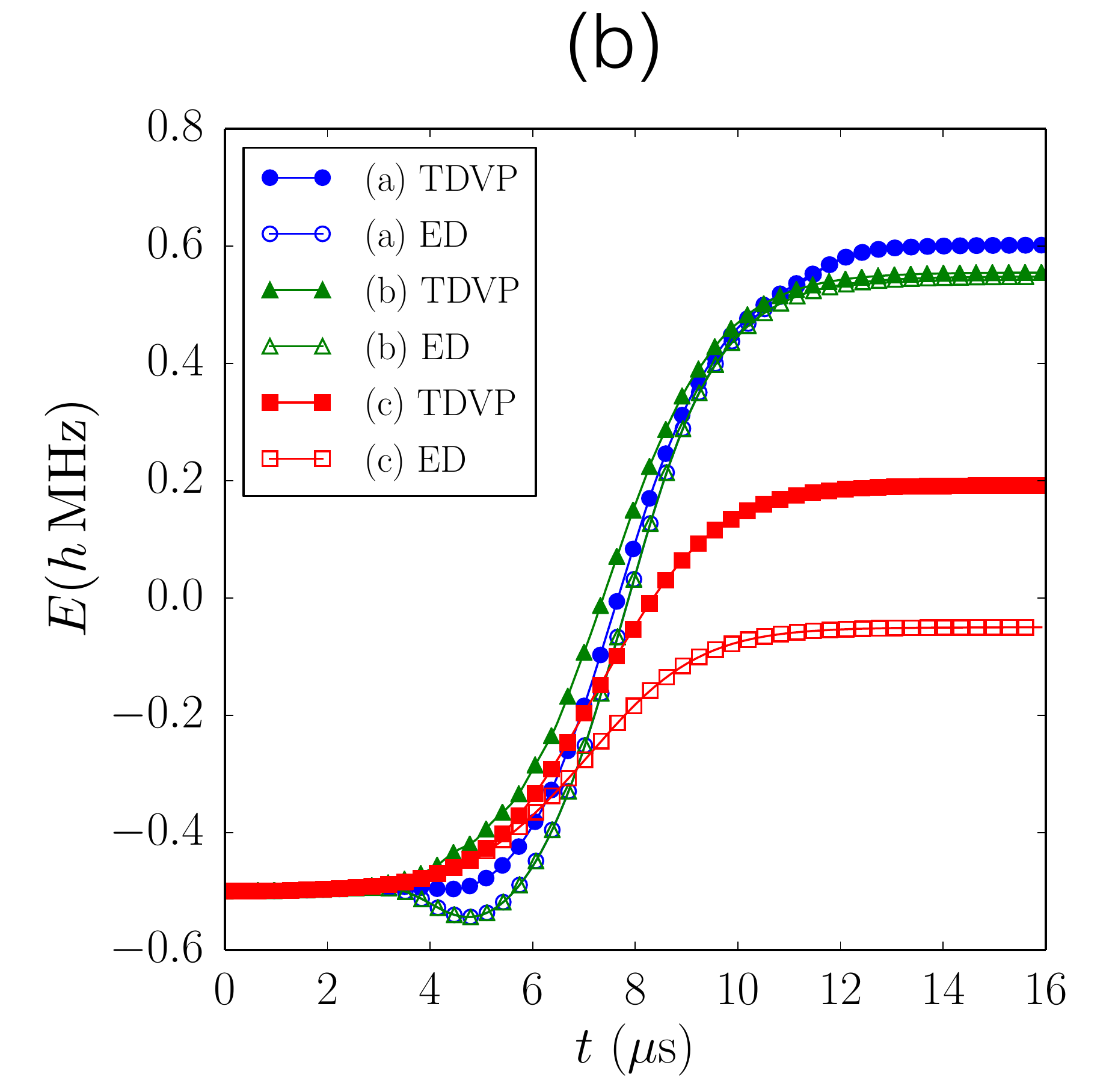}
\includegraphics[height=4cm]{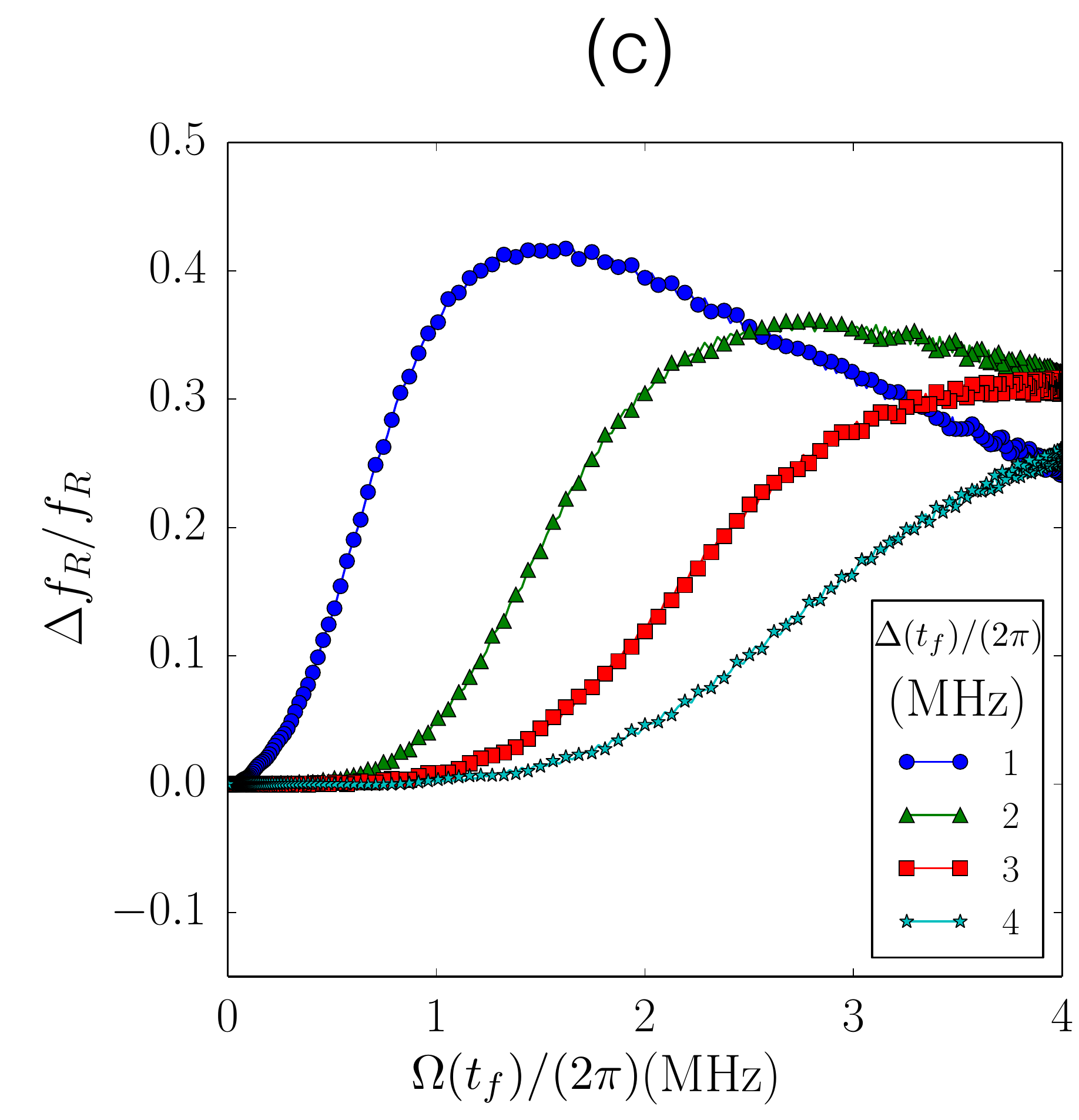}
\protect\caption{Comparison between TDVP and ED. (a) density of Rydberg atoms
$f_{R}$ as a function time $t$ during the sweep for the three paths
shown in \fref{fig:sketch}. (b) Energy as function of
sweep time. (c) graph shows the difference in the Rydberg density
$\Delta f_{R}/f_{R}$ after an adiabatic parameter sweep as calculated
from TDVP with respect to ED as a function of the final $\Omega(t_f)/(2\pi)$
and for four different final detunings $\Delta(t_f)/(2\pi)=1,2,3,4$ MHz.
For $\Delta(t_f)/(2\pi)=2$ MHz the difference $\Delta f_{R}$ of the Rydberg
density between the two methods starts to deviate from zero if $\Omega(t_f)/(2\pi)\gtrsim1$ MHz,
indicating the breakdown of our variational approach. \label{fig:comparvst}}
\end{centering}
\end{figure}

\Fref{comparison} shows final distributions of Rydberg atoms
computed using our TDVP approach (upper panel) as well as ED (lower
panel) for the three sweeps (a) (b) (c). Note that in contrast to the 1D case, the variational ansatz
describes the sweep to the classical limit $\Omega(t_f)=0$ perfectly
well, even though our approach propagates the wave-function through
the non-classical region $\Omega>0$, where our ansatz is not strictly
valid. In a perfectly adiabatic situation corresponding to $t_f\to\infty$, the final state obtained with our time-dependent variational approach would coincide with the variational ground state, which is the exact ground state in this case. The fact that our results for a finite sweep time compare very well with the exact solution suggests that deviations from adiabaticity are negligible. For such a small system size, the competition between the laser excitation and the vdW interactions results in a regular pattern of four Rydberg atoms placed at the corners of the system. We also obtain a good agreement for $\Omega(t_f)/(2\pi)=1$ MHz and the corresponding pattern is not modified compared to the classical limit. However,
for $\Omega(t_f)/(2\pi)=4$ MHz, our ansatz overestimates the ground state energy considerably, even though
the distribution of excitations looks similar to the exact result.
It is also instructive to study how the system behaves during the sweep. 
\Fref{fig:comparvst}s (a) and (b) show a comparison of Rydberg density
$f_{R}$ and energy as function of time during the parameter sweep
for the three sweep protocols shown in \fref{fig:sketch}. Again
a substantial difference between TDVP and ED is only visible for sweeps
to large final Rabi frequencies.

The results shown in \fref{comparison} and \fref{fig:comparvst}(b),(c) allow to assess the validity of our approach for a realistic dynamical state preparation. We are also interested in estimating the typical value of the parameters $\Omega$, $\Delta$ where our ansatz can describe the ground state of the model \eref{eq:H}, regardless of the details of the dynamical state preparation. To this end we consider a very large sweep time $t_f=150\ \mu$s to ensure that the equations of motion \eref{eq:eq_motion} lead to the formation of the variational ground state whereas the solution obtained with ED results in the exact ground state. We then estimate the regime of validity of the variational ground state as follows: we compute the Rydberg density at the end of the parameter sweep using both TDVP and ED and plot the relative difference
of $f_{R}$ between the two approaches as a function of the final
Rabi frequency $\Omega(t_f)$ at the end of the sweep. Results are shown in \fref{fig:comparvst}(c) 
for four different final detunings $\Delta$. We see that for a final detuning $\Delta(t_f)/(2\pi)=2$
MHz the difference $\Delta f_{R}/f_{R}$ starts to deviate from zero
if the sweep protocol samples Rabi frequencies which are larger than
$\Omega(t_f)/(2\pi)>1$ MHz. Accordingly, for $\Delta/(2\pi)=2$ MHz our variational
ansatz is correct as long as $\Omega/(2\pi)\lesssim1$ MHz. We note, however, that this criterion was obtained for a small system and potentially depends on system size.

\subsection{Isotropic Rydberg crystals}
Now that we have assessed the regime of validity of our ansatz and checked in particular that it can quantitatively describe the dynamical preparation of Rydberg crystals in small two dimensional systems, let us now present our results for large system sizes
where an exact numerical treatment is not possible. 

We first describe the formation of Rydberg crystals in an isotropic configuration. In analogy to the experimental setup \cite{Schauss2014}, we start from a circular (cookie
shaped) distribution of $N=777$ ground-state atoms considering the three sweeps path shown in \fref{fig:sketch} keeping the other
parameters of the last subsection unchanged.

\begin{figure}
\begin{centering}
\includegraphics[width=0.6\columnwidth]{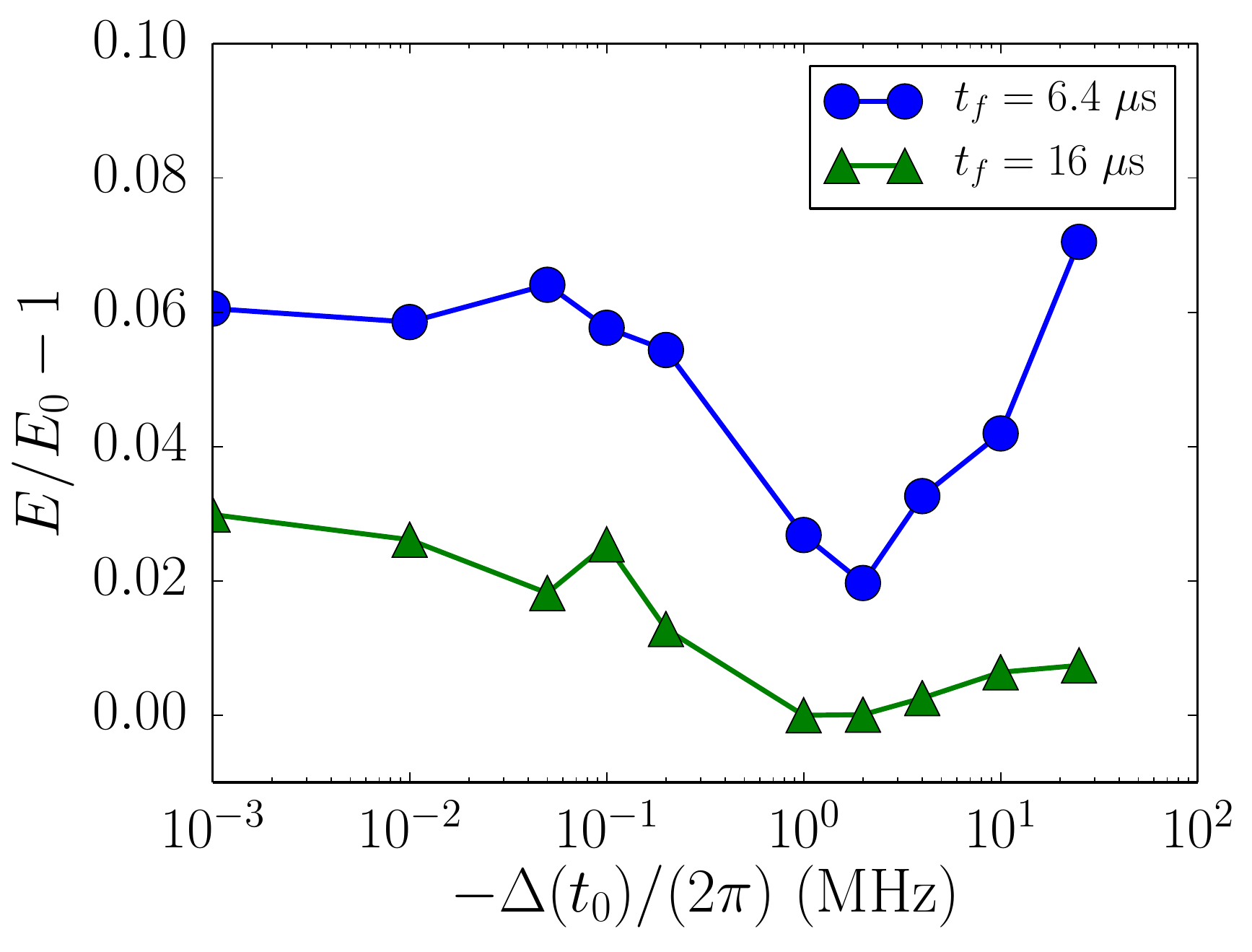} 
\protect\caption{Influence of the sweep time $t_{f}$ and of the initial detuning
$\Delta(t_0)$ on the energy $E$ at the end of the sweep for the case
of isotropic interactions and  $\Omega(t_f)=0$ and $\Delta(t_f)/(2\pi)= 2$ MHz.
$E_{0}$ represents the minimum of the energy obtained for $t_{f}=16\ \mu$s
and $\Delta(t_0)/(2\pi)$= -1 MHz which corresponds to the first path in \fref{fig:sketch}. The results show that the fidelity
to stay in the ground-state during the parameter sweep decreases if
$\Delta(t_0)$ is too small or the sweep time is too short.\label{fig:Delta0}}
\end{centering}
\end{figure}

Let us first comment on our choice of sweep paths (\fref{fig:sketch}). In order to prepare a state which is as close as possible to the variational ground-state, it is particularly important
to circumvent the region around the critical point $\Omega=0,\Delta=0$
\cite{Weimer2008} during the sweep into the crystalline phase. Indeed we
found that the energy of the final state increases substantially if
the initial negative detuning is chosen too small. On the other hand,
if the initial detuning is too large, the length of the sweep path
in parameter space is long and the rate of change of the parameters
during the same sweep time is increasing such that the sweep becomes
less adiabatic again. This is shown in \fref{fig:Delta0} where
we plot the energy of the final state obtained for $\Omega(t_f)=0$ and $\Delta(t_f)/(2\pi)= 2$ MHz as a function of the initial
detuning, for different sweep times. We found that the optimal
choice of the initial detuning is $\Delta(t_0)/(2\pi)=-1$ MHz. In this
case our sweeps are almost adiabatic in the sense that the energy
of the states at the end of the sweeps is less than three percent
above the ground-state energy which we found by an independent optimization
of the variational wave-functions using a homotopy-continuation method
\cite{Punk2014}.

\begin{figure}
\begin{centering}
\includegraphics[width=0.30\columnwidth]{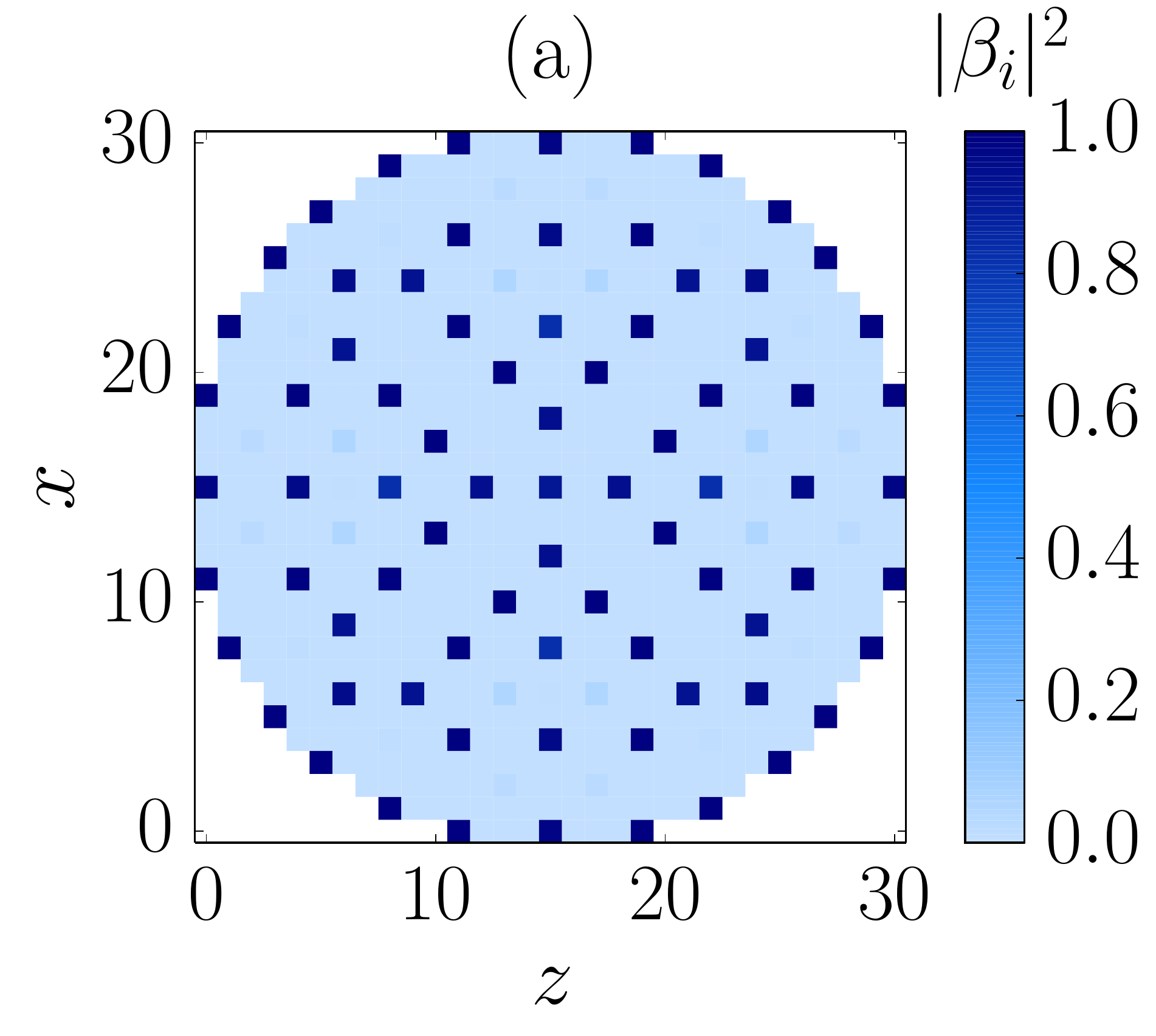}
\includegraphics[width=0.30\columnwidth]{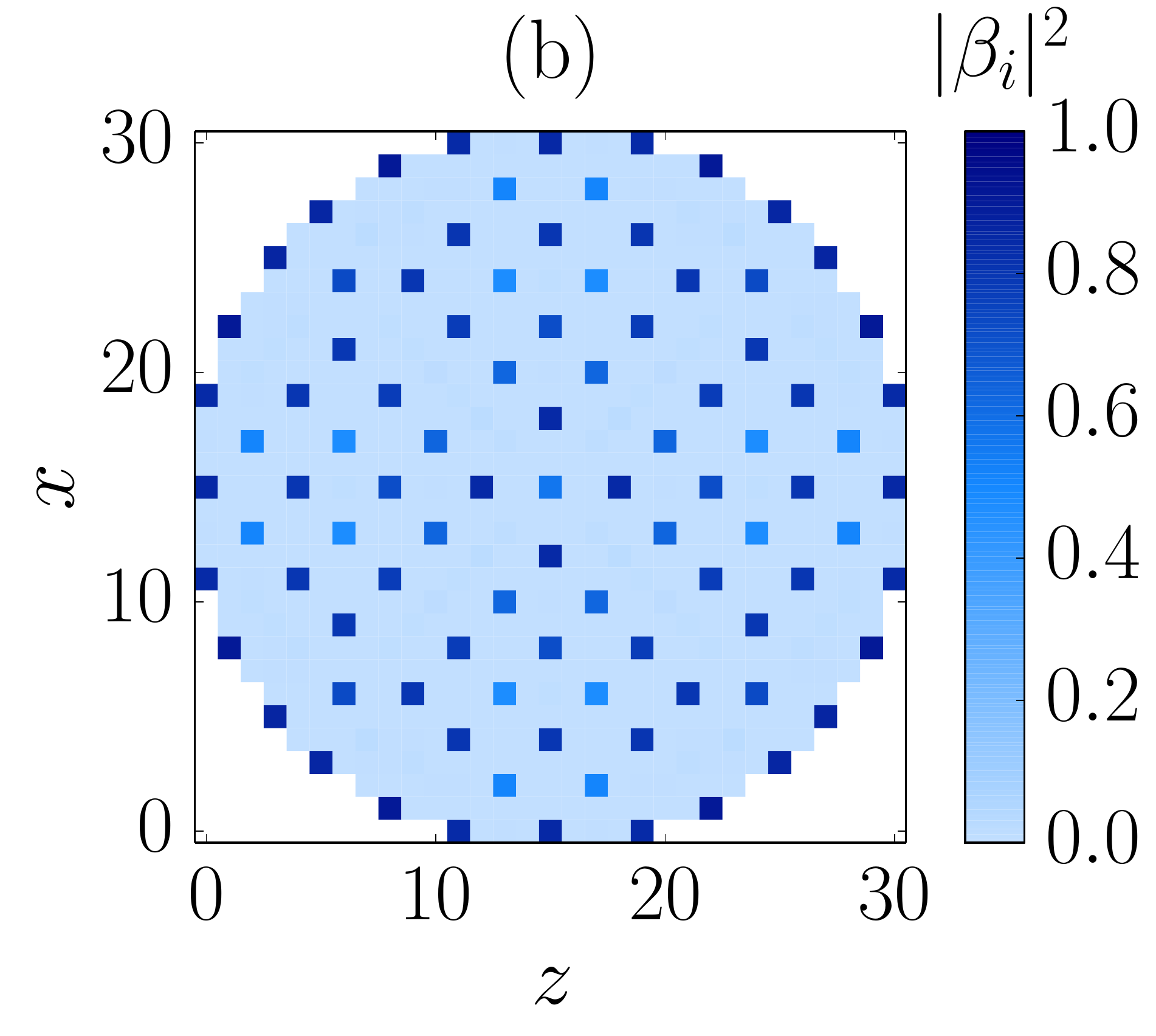}
\includegraphics[width=0.30\columnwidth]{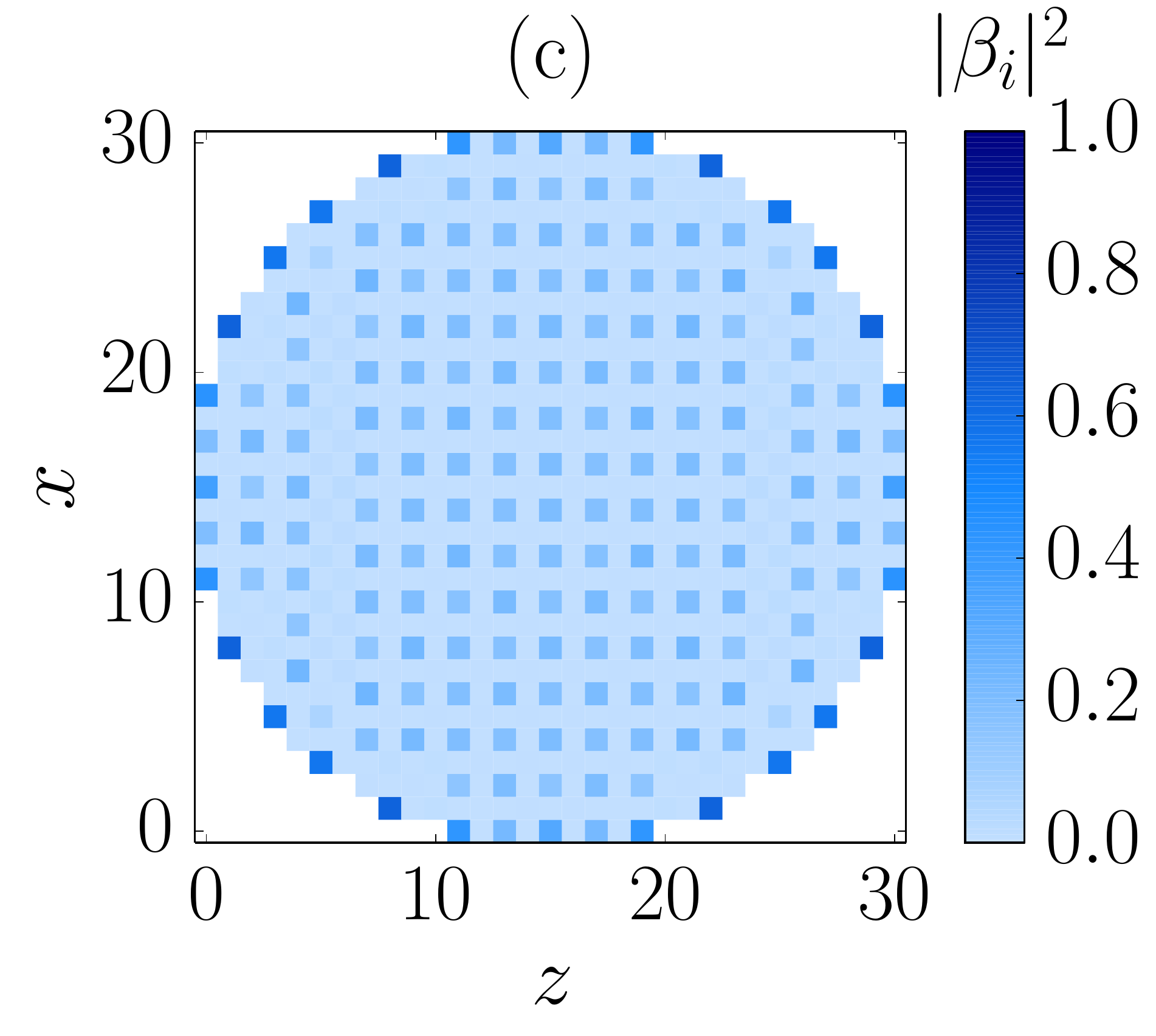}
\protect\caption{Distribution of Rydberg excitations at the end of the three parameter
sweep protocols (a), (b), (c) shown in \fref{fig:sketch} for
the case of isotropic interactions. Plotted is the probability $|\beta_{i}|^{2}$
to be in the Rydberg state on each lattice site. The corresponding
energies per lattice site are $E=0.86,0.82,0.52\,h\,$MHz. The crystalline
arrangement of Rydberg atoms is clearly visible. In the classical
limit $\Omega\to0$ (left) the excitation probabilities are close
to either zero or one, whereas quantum superpositions with intermediate
values of $|\beta_{i}|^{2}$ appear at finite $\Omega$ (middle, right).
\label{fig:iso}}
\end{centering}
\end{figure}

Results for the final distribution of Rydberg excitations at the end
of the sweep are shown in \fref{fig:iso}. We note that for all three sweep protocols we obtain a single crystalline
pattern which respects the symmetries of the cookie-shaped atom distribution.
The shape of the crystal is pinned by boundary effects and the variational
ground-state that we find is non-degenerate and unique. This is in contrast to
experiments, where different
symmetry-related, almost degenerate crystal configurations are observed
from shot to shot~\cite{Schauss2014}. Also note that our equations
of motion for the variational parameters \eref{eq:eq_motion} preserve
symmetries during time evolution. If degenerate, symmetry related
ground states exist for a given set of parameters $\Delta$ and $\Omega$,
the time evolution passes through a bifurcation point, which signals
the presence of the phase transition to the crystalline state. At
this point tiny numerical errors will pick out one of the degenerate
ground states. It is important to emphasize, however, that we always
found a unique, symmetric variational ground-state for the parameters
considered here.

Ideally, the first sweep to a final Rabi frequency $\Omega(t_f)=0$ shown in \fref{fig:sketch}(a) prepares
the ground-state of the classical Ising model, if the sweep were perfectly adiabatic. In this case the arrangement of excited Rydberg
atoms would correspond to the minimum energy configuration of classical
charges with a $1/r^{6}$ potential and the probability to be in the
Rydberg state is either zero or one in this limit. From \fref{fig:iso}(a)
one can see that the probability to be in the Rydberg state is $\sim0.8$
rather than $0$ or $1$ on some sites, indicating deviations from
adiabaticity. Nevertheless, a crystalline arrangement of Rydberg atoms
is clearly visible. The average density of Rydberg atoms is $f_R=0.09$
in this case, which is in accordance with an average distance between
two excitations on the order of $\left[C_{6}\left(\frac{\pi}{2}\right)/\hbar\Delta\right]^{1/6}\approx2-3$.

For the case of sweeps to finite final Rabi frequencies $\Omega(t_f)$
away from the classical limit [figures~\ref{fig:iso}(b) and (c)], quantum superpositions between ground-state
and excited atoms are present, and the probability to be in the Rydberg
state is thus no longer restricted to zero or one. For increasing
final $\Omega$ quantum fluctuations are stronger and the average
number of Rydberg excitations increases, while the average excitation
probability decreases. At large enough $\Omega$ the crystalline arrangement
finally melts and one enters a quantum disordered regime where the
average excitation probability is equal on all lattice sites. This
trend is visible in panel (c). 

Note that the complex crystalline arrangement of Rydberg atoms is
strongly dependent on the size and of the shape of the system. In
an infinite system the excited atoms would ideally maximize their
average distance, which would result in a triangular lattice of Rydberg
atoms. Due to the underlying square optical lattice strong commensurability
issues arise, however, in particular if the average distance between
excitations is on the order of a few lattice spacings. We observe
that the crystalline structure is strongly pinned by boundary effects
in our case and the crystalline structures in the classical limit
thus do not resemble those which supposedly exist in the thermodynamic
limit \cite{Rademaker2013}.

\subsection{Anisotropic Rydberg crystals}

We now describe the preparation of anisotropic Rydberg crystals. In this case, the magnetic field is set along
the $z$-direction of the optical lattice, as shown in \fref{fig:setup}(a),
keeping the other parameters such as sweep paths, atom distribution and the Rydberg level unchanged.
Accordingly, the interaction between Rydberg atoms is anisotropic
and stronger in x- than in z-direction.

\begin{figure}
\begin{centering}
\includegraphics[width=0.30\columnwidth]{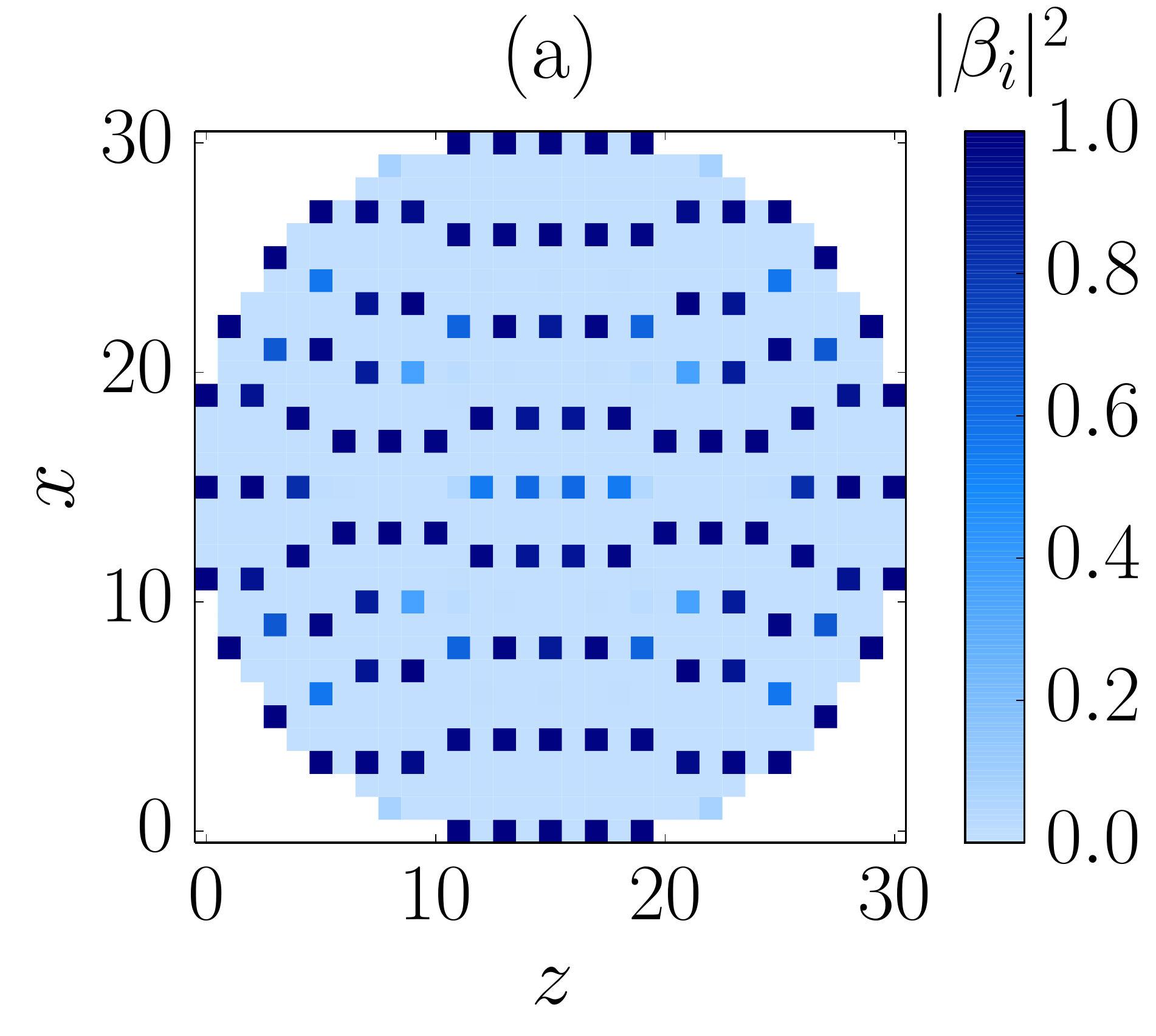}
\includegraphics[width=0.30\columnwidth]{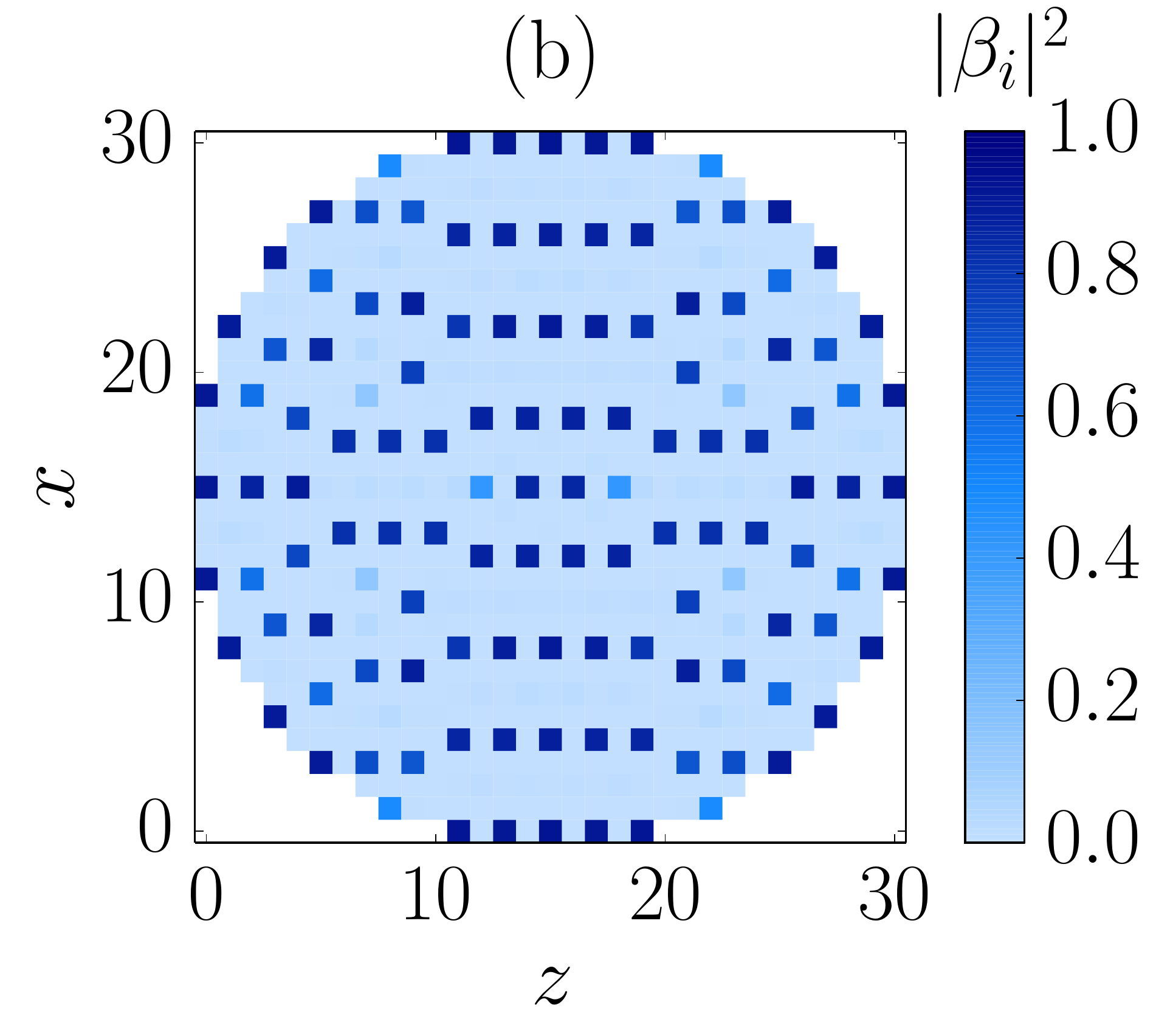}
\includegraphics[width=0.30\columnwidth]{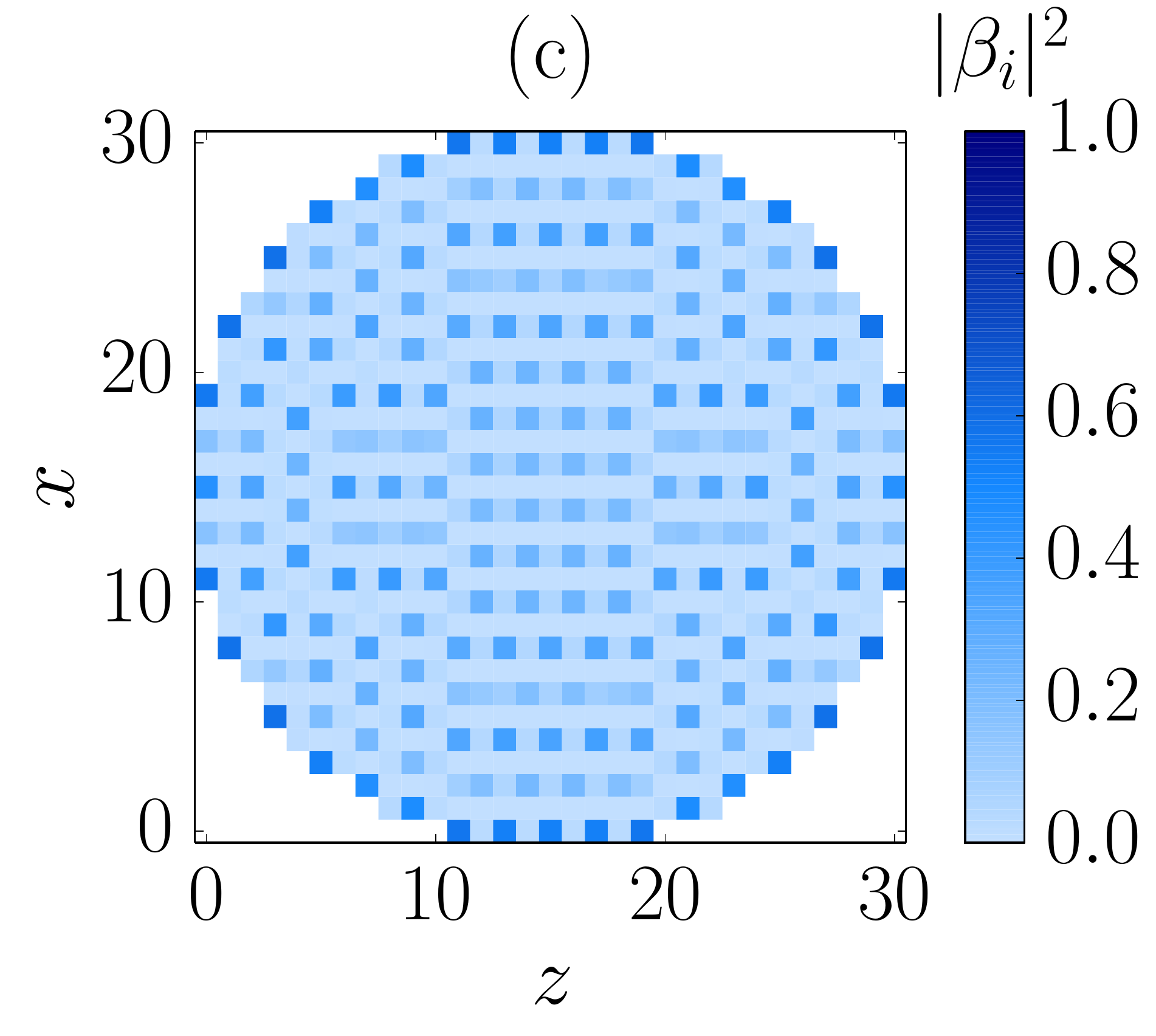}
\includegraphics[width=0.30\columnwidth]{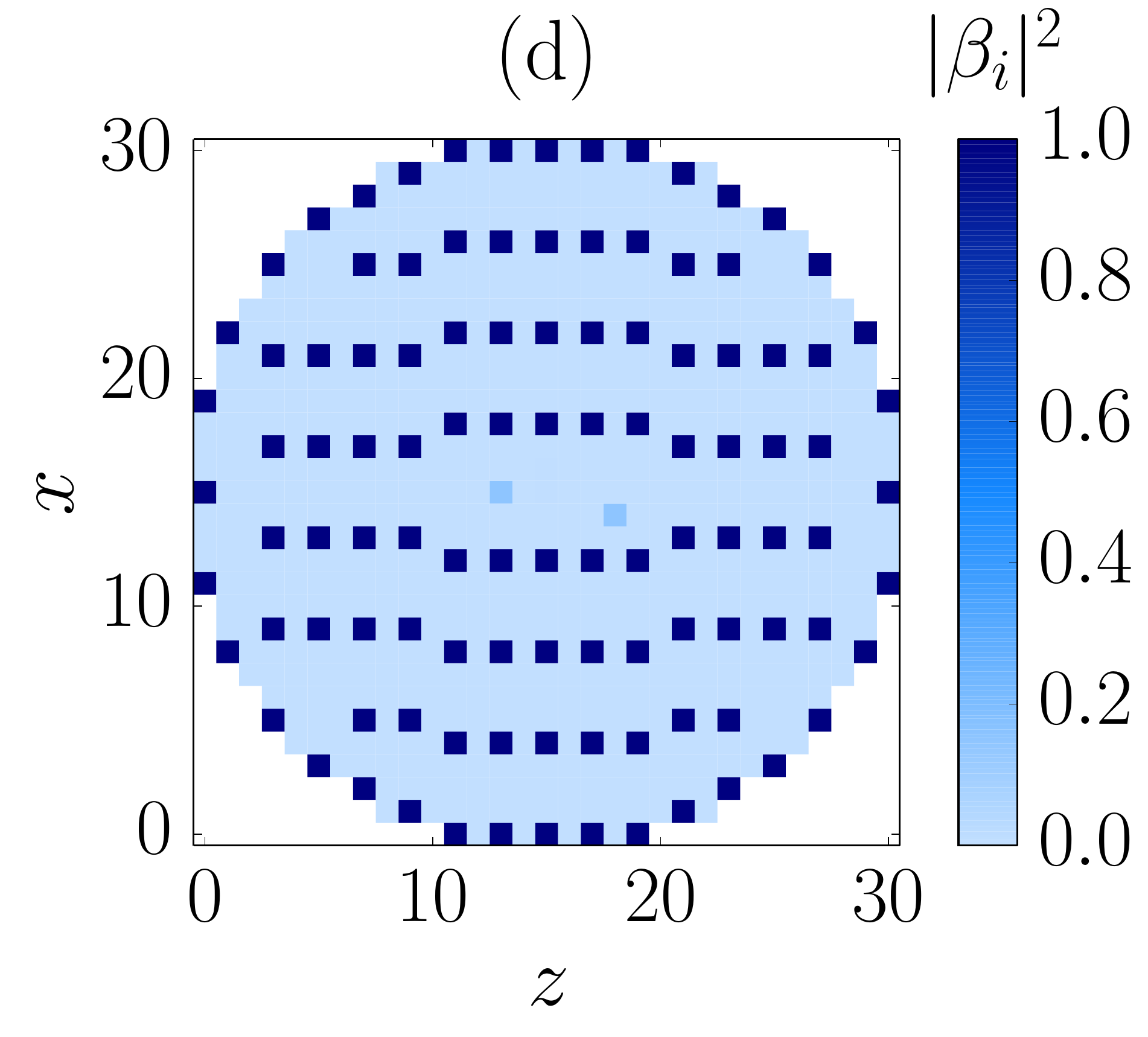}
\includegraphics[width=0.30\columnwidth]{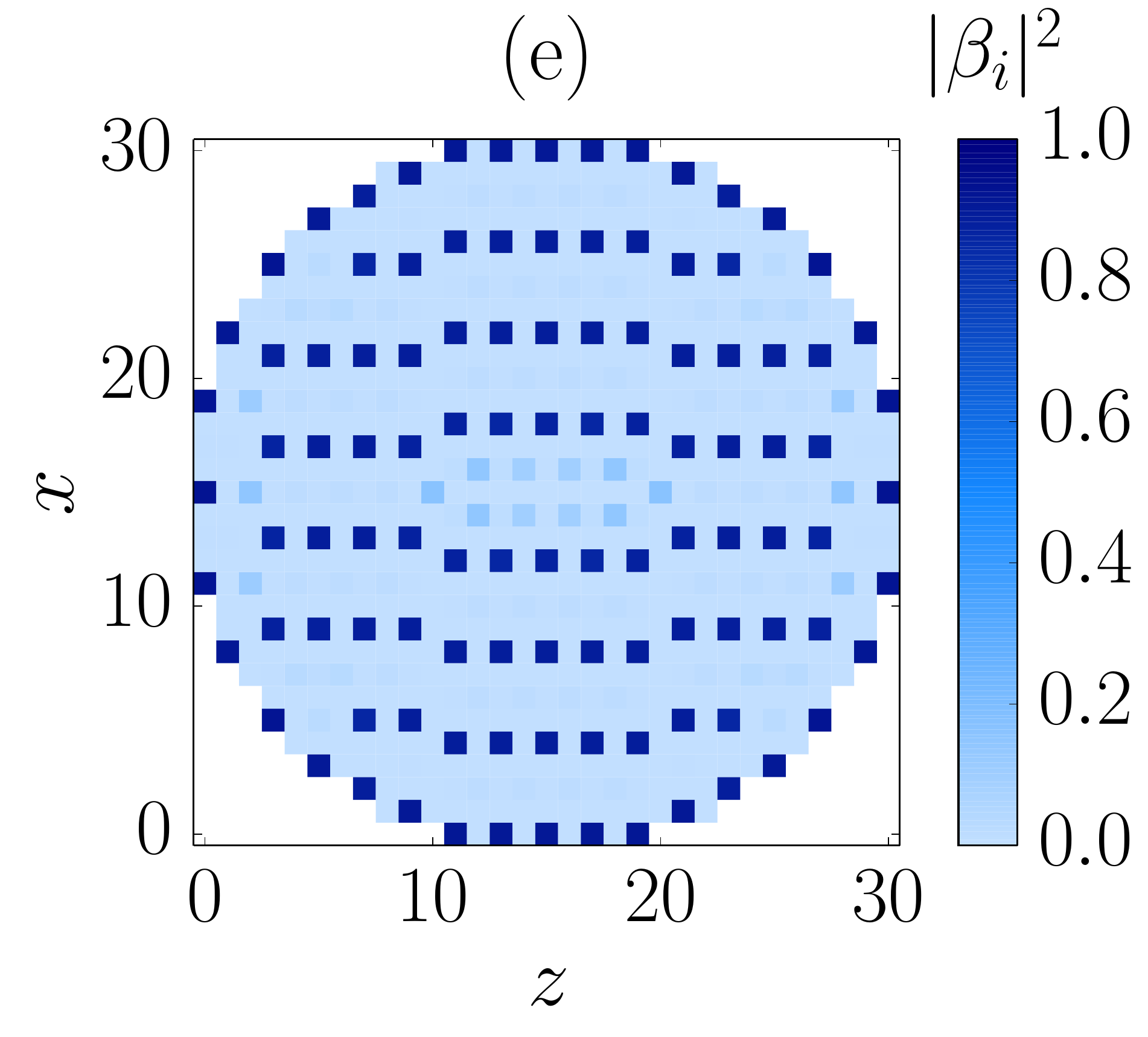}
\includegraphics[width=0.30\columnwidth]{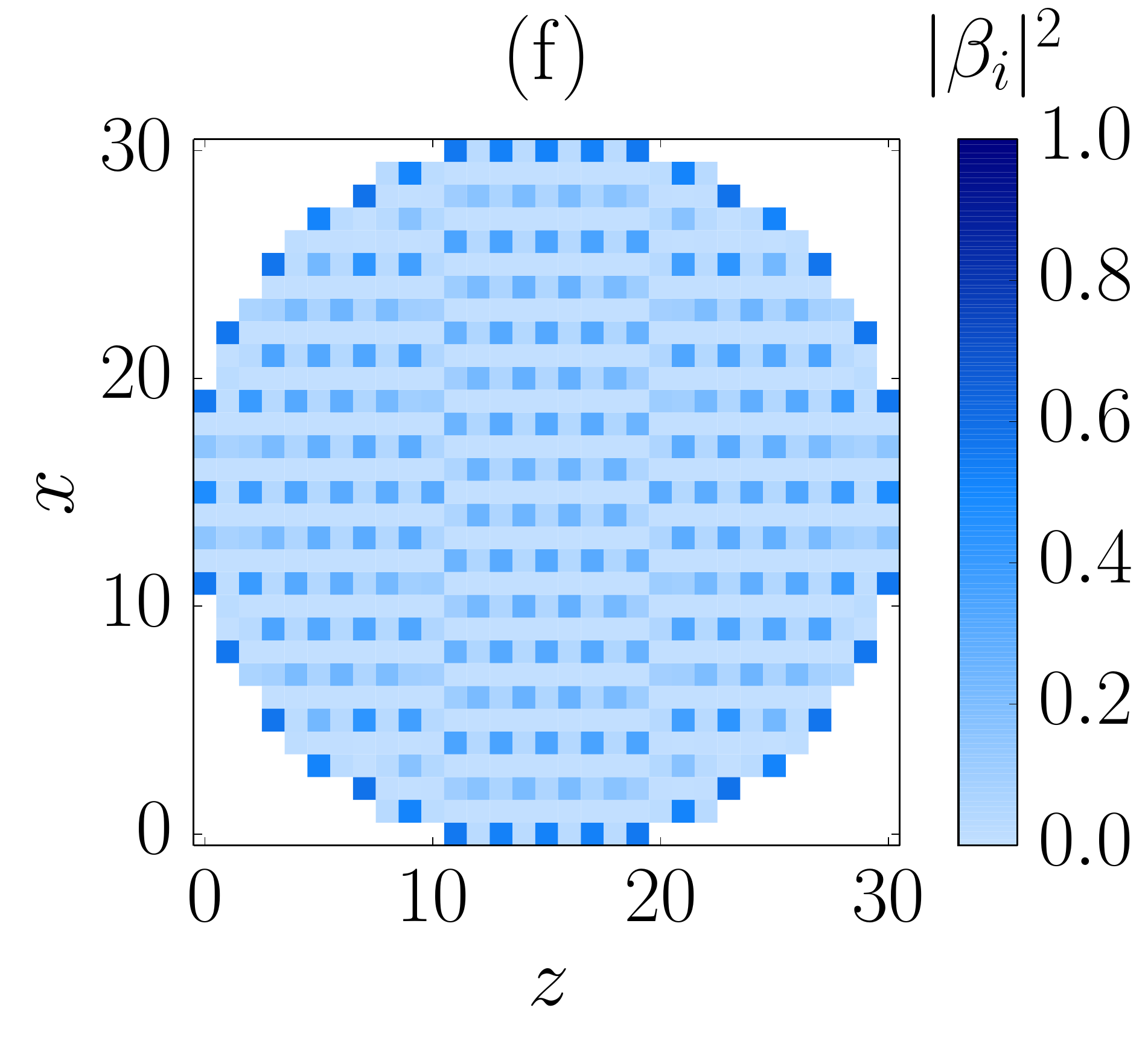}
\protect\caption{Upper panel: distribution of Rydberg excitations at the end of the
three parameter sweep protocols (a), (b), (c) shown in \fref{fig:sketch}
for the case of anisotropic interactions. Corresponding energies per
lattice site are $E=0.79,0.74,0.30\,h $ MHz. Lower panel: distribution
of Rydberg excitations obtained after a direct minimization of the
variational ground-state energy. The Rabi frequencies and detunings
match the parameters at the end of the sweep protocols in the upper
panel. Corresponding ground-state energies per particle are $E=0.77,0.72,0.29\,h\,$
MHz. Defects in the crystalline arrangement due to the small non-adiabaticity
of the sweep protocols are clearly visible in the left panel. \label{fig:aniso}}
\end{centering}
\end{figure}

Results for the three sweep paths are presented in the upper panel
of \fref{fig:aniso}. As in the isotropic case, the crystal progressively
melts as $\Omega$ is increased. Note, however, that the anisotropy
is visible in all cases and the crystalline structure melts first
in the weakly interacting z-direction while translational symmetry
is still broken in the strongly interacting x-direction. Again, we observe that the form of the Rydberg crystal
is strongly pinned by boundary effects, similar to the isotropic case.

In the classical limit $\Omega=0$, we find an anisotropic crystal with an average distance between excitations on the order of $\left[C_6\left(\frac{\pi}{2}\right)/\hbar\Delta\right]^{1/6}\approx 3$ in the $x-$direction and of $\left[C_6\left(0\right)/\hbar\Delta\right]^{1/6}\approx 1-2$ in the $z-$direction. 
Again, the results for the sweep to the classical limit $\Omega(t_f)=0$ 
indicate that the preparation was not perfectly adiabatic.
Indeed, the excitation probabilities differ from 0 or 1 at the end of the parameter sweep, as in the case of isotropic interactions. The deviations from adiabaticity are even more pronounced for anisotropic interactions, as we
find non-classical excitation probabilities on the order of $\sim0.5$
in this case. This can be attributed to the fact that the excitation gaps are smaller compared to the case of isotropic interactions, due to the substantially weaker interaction in z-direction.

In order to estimate the fidelity of the dynamic state preparation scheme we plot  the distribution
of Rydberg atoms obtained after a direct minimization of the ground-state
energy within our variational ansatz in the lower panel of \fref{fig:aniso}. Comparing this to the distributions
obtained after the parameter sweep it is apparent that a number of
defects are created due to the not fully adiabatic sweep protocol. Again, the
crystalline arrangement is not strongly affected by the rather short sweep
time, however.

\section{Conclusion and outlook}

In the present work we have developed a time dependent mean field theory to model the dynamical preparation of anisotropic Rydberg crystals with atoms in 2D optical lattices. In addition we have presented results of numerical simulations relevant for experimentally realistic system sizes, in the limit of patterns with a large number of Rydberg excitations.

We note that the anisotropic character of the vdW interactions has been seen experimentally in a recent Rydberg-blockade experiment involving Rydberg s and d-states \cite{Barredo2014}. In contrast to the present work, where we considered the anisotropic vdW interactions between single Zeeman levels of the Rydberg states, i.e. in the limit of large Zeeman splitting, in these experiments vdW couplings involving transitions between Zeeman levels can be important. This interplay leads to several new physical phenomena, which will be presented in a future work. 
\ack
We thank W. Lechner, T. Lahaye, A. Browaeys and I. Bloch for interesting discussions. Work at Innsbruck is supported
by the ERC-Synergy Grant UQUAM and SFB FOQUS of the Austrian Science Fund. B. V. acknowledges the Marie Curie
Initial Training Network COHERENCE for financial support. M. P. is
supported by the Erwin Schrödinger Fellowship No. J 3077-N16 of the
Austrian Science Fund (FWF).

\section*{References}
\bibliographystyle{iopart-num}
\bibliography{Crystal}

\end{document}